\documentclass[11pt]{article}

\usepackage[margin=1in]{geometry}
\usepackage{amsmath,amssymb}
\usepackage{booktabs}
\usepackage{graphicx}
\usepackage{xcolor}
\usepackage{hyperref}
\usepackage{braket}
\usepackage{authblk}
\usepackage{tabularx}
\usepackage{array}

\title{Decomposing Gradient Suppression in Barren Plateaus: Activity, Sign Organization, and Coupling}
\author{Pilsung Kang}
\affil{Department of Software Science, Dankook University, Yongin, 16890, South Korea\\ pilsungk@dankook.ac.kr}
\date{}                     
\begin{document}
\maketitle

\begin{abstract}
Barren plateaus (BPs) are conventionally characterized by the suppression of
gradient variance, but this aggregate description does not reveal how the loss
of gradient signal is composed across individual Hamiltonian terms. We introduce
a term-resolved diagnostic framework that decomposes the second moment of the
gradient exactly into three factors: pre-cancellation activity, measuring the
overall magnitude of termwise contributions before signed cancellation; sign
organization, measuring how their signs reinforce or cancel in the aggregate
gradient; and coupling, measuring the statistical association between activity
and organization. As a reference for unstructured cancellation, we use a
conditional random-sign model that preserves the termwise magnitudes while
treating their signs as independent and symmetric, yielding exact reference
predictions for organization and coupling while leaving activity unconstrained.
We apply the decomposition to a hardware-efficient ansatz (HEA) and a
Hamiltonian variational ansatz (HVA) for the transverse-field Ising model (TFIM)
and the longitudinal-field Ising model (LFIM). For the HEA, the finite-size
suppression of the second moment is carried almost entirely by decaying termwise
activity, accounting for $96.5$--$98.9\%$ of the fitted log-slope across the
tested depths in both Hamiltonians, while organization shows no systematic
scaling and coupling remains consistent with its random-sign reference. For the
HVA, activity and organization instead grow with system size and contribute
comparably to the second-moment scaling. A bias-corrected mean-gradient check is
consistent with zero in every tested condition, so the second-moment results
carry over approximately to the gradient variance. A microscopic sign-alignment
analysis further shows sector-structured organization in the HVA, whereas the
HEA exhibits only weak residual sign structure that does not accumulate into net
signed organization. These patterns are reproduced across both Hamiltonians. The
decomposition thereby provides a term-resolved characterization of internal
gradient-scaling composition that is not apparent from aggregate gradient
statistics alone in the regimes studied.

\end{abstract}

\section{Introduction}
\label{s:intro}

Variational quantum circuits have emerged as one of the central algorithmic
paradigms for near-term quantum computing, but their practical usefulness is
often limited by trainability~\cite{peruzzo:2014:vqe,farhi:2014:qaoa,cerezo:2021:variational}. 
A particularly severe obstacle is the barren plateau (BP) phenomenon, in which the
optimization landscape becomes exponentially flat as system size grows, causing
gradient-based training to lose usable signal and rendering variational learning
increasingly difficult~\cite{mcclean:2018:barren}. Since its original
identification in random parameterized quantum circuits, BP has been recognized
as a central bottleneck in variational quantum algorithms, and subsequent work
has shown that its onset is closely tied to circuit expressibility,
concentration of measure, and the suppression of gradient variance in
high-dimensional Hilbert
space~\cite{holmes:2022:prx,cerezo:2021:natcomm,larocca:2025:natreview}. 

The standard theoretical account of BP is therefore formulated in terms of
gradient concentration: for sufficiently expressive or effectively random
parameterized circuits, the variance of cost-function gradients decreases
rapidly with system size, often exponentially, so that optimization becomes
increasingly dominated by noise rather than usable learning
signal~\cite{wang:2021:natcomm}. 
While this perspective provides a powerful asymptotic characterization
of trainability, it does not by itself resolve how the aggregate
gradient statistics are composed at the level of individual
Hamiltonian-term contributions.
In particular, when the cost
function gradient $\partial_k\langle H\rangle$ is written as a sum of termwise
contributions, the standard formulation does not explicitly distinguish whether
vanishing gradients arise because only a few terms remain active, because many
terms contribute with small magnitude, or because substantial positive and
negative contributions cancel one another. 
This unresolved termwise composition motivates the present work.

In this work, we analyze gradient suppression by resolving each
gradient into signed contributions from individual Hamiltonian
terms. This termwise view distinguishes the magnitude of
pre-cancellation activity from the organization of contribution
signs. When the gradient $\partial_k\langle H\rangle$ is decomposed
in this way, a small aggregate signal can reflect weak termwise
activity, signed cancellation among active contributions, or a
combination of the two. The value of this viewpoint is that these
components of gradient suppression become separately measurable.
Whether the system-size dependence of the gradient signal in a
given circuit family is carried by changing termwise activity,
changing sign organization, or coupling between the two can then be
posed as an empirical question, which this paper addresses for two
contrasting ansatz families.

Recent work on BPs has approached trainability from several complementary
angles, including diagnoses of problem-inspired ansatzes using tools from
quantum optimal control~\cite{larocca:2022:quantum}, adjoint-representation and
Lie-algebraic analyses of gradient concentration and circuit
structure~\cite{fontana:2024:natcomm,ragone:2024:natcomm}, architecture-level
analyses based on alternative formalisms such as ZX-calculus~\cite{zhao:2021:quantum}, 
and optimization-side mitigation through adaptive, nested, or warm-start variational
strategies~\cite{li:2026:npj,puig:2025:prx,zunkovic:2026:adiabatic}. 
In contrast, our goal is not to propose another training heuristic,
but to provide a diagnostic decomposition of the gradient second
moment into termwise activity, sign organization, and their
coupling, connected to the standard variance-based description
through an exact second-moment identity, with the step from the
second moment to the variance checked empirically.

Our contributions are threefold:
\begin{itemize}

\item We introduce an exact decomposition of the second moment of the
gradient into pre-cancellation activity, sign organization, and their
statistical coupling, built on the termwise diagnostics of the
cancellation ratio $R_k$, the effective term count
$N_{\mathrm{eff},k}$, and the organization diagnostic
$B_{\mathrm{eff},k}=R_k\sqrt{N_{\mathrm{eff},k}}$. Under a
conditional random-sign model that preserves contribution magnitudes
while treating their signs as independent and symmetric, the
organization and coupling factors obey exact reference predictions.
We further connect the second-moment decomposition to the standard
gradient variance through the exact variance identity and a
bias-corrected mean-gradient check, which is consistent with zero
across all tested conditions.

\item We apply the decomposition to a hardware-efficient ansatz
(HEA)~\cite{kandala:2017:hea} and a Hamiltonian variational ansatz
(HVA)~\cite{wecker:2015:hva} for the transverse-field Ising model
(TFIM) and the longitudinal-field Ising model (LFIM). For the HEA,
the finite-size suppression of the second moment is carried almost
entirely by decaying termwise activity, which accounts for
$96.5$--$98.9\%$ of the fitted log-slope across the tested depths in
both Hamiltonians, while organization remains near the random-sign
reference without systematic scaling and coupling remains
statistically consistent with its reference. In contrast, for the
HVA, activity and organization both grow with system size and
contribute comparably to the second-moment scaling, while the
coupling factor provides the clearest quantitative difference
between the two Hamiltonians.

\item We characterize the microscopic structure underlying the
organization factor through a sign-alignment analysis. The HVA
exhibits pronounced within-sector alignment associated with
Hamiltonian sectors, whereas the HEA shows only weak residual sign
structure that does not accumulate into net signed organization.
This analysis provides a microscopic counterpart to the distinct
aggregate organization profiles identified by the decomposition.

\end{itemize}

The remainder of the paper is organized as follows.
Section~\ref{s:framework} introduces the termwise diagnostics and
the conditional random-sign model. Section~\ref{s:var_bridge}
derives the exact second-moment decomposition and its two
falsifiable predictions. Section~\ref{s:decomposition} presents the
measured decomposition for both ansatz families and both
Hamiltonians, and Section~\ref{s:microscopic} characterizes the
underlying sign-alignment structure. Section~\ref{s:discuss}
discusses scope and limitations, and Section~\ref{s:conc}
concludes. Methods, including all estimands, estimation procedures,
and the sign-alignment analysis, are collected in
Section~\ref{s:methods}.

\section{A Termwise Diagnostic Framework for Signed Cancellation}
\label{s:framework}

In this section, we introduce the termwise diagnostic framework on
which the rest of the paper is built. We first decompose each
gradient into contributions from individual Hamiltonian terms, then
define complementary diagnostics of signed cancellation, and
finally introduce the conditional random-sign model that provides
aggregate reference predictions for unstructured cancellation.

\subsection{Termwise Gradient Contributions}

We begin by decomposing the problem Hamiltonian into Pauli terms as
\begin{equation}
H = \sum_{\alpha} c_{\alpha} P_{\alpha},
\end{equation}
where $\alpha$ indexes the individual Pauli terms in the Hamiltonian,
$P_{\alpha}$ is the corresponding Pauli string, and $c_{\alpha}\in\mathbb{R}$ is
its corresponding coefficient. This decomposition is standard in variational
quantum algorithms, but here it plays a more specific conceptual role: it allows
the gradient of the cost function to be resolved into contributions associated
with individual Hamiltonian terms rather than treated only as a single aggregate
quantity.  For a variational state $\ket{\psi(\theta)}$, we define the cost
function as
 \begin{equation}\label{eq:cost}
C(\theta)
=
\langle H\rangle_{\theta}
=
\bra{\psi(\theta)}H\ket{\psi(\theta)}.
\end{equation}
The gradient with respect to parameter $\theta_k$ can therefore be written as
\begin{equation}
\partial_k \langle H\rangle
=
\sum_{\alpha} c_{\alpha}\,\partial_k \langle P_{\alpha}\rangle.
\end{equation}
This termwise resolution is the starting point of our framework, because it
makes explicit that a small total gradient need not imply that all underlying
contributions are individually small. Instead, a vanishing gradient may also
arise when many nontrivial termwise contributions carry competing signs and
cancel one another.

To make this decomposition operational, we define the \emph{termwise gradient
contribution} of Hamiltonian term $\alpha$ to parameter $\theta_k$ as
\begin{equation}
a_{\alpha,k}
\;:=\;
c_{\alpha}\,\partial_k \langle P_{\alpha}\rangle.
\end{equation}
With this notation, the full gradient of the cost function is written compactly as
\begin{equation}\label{eq:termwise_grad}
\partial_k \langle H\rangle
\;=\;
\sum_{\alpha} a_{\alpha,k}.
\end{equation}
This definition separates the gradient into a collection of signed contributions
indexed by Hamiltonian terms, and it is precisely this signed structure that our
framework aims to analyze.  In particular, the central object is no longer only
the magnitude of the aggregate gradient $\partial_k \langle H\rangle$, but also
the internal organization of the termwise contributions $a_{\alpha,k}$ for a
fixed parameter $k$: whether these contributions are mutually aligned, weak but
sign-aligned, or large in magnitude yet strongly cancelling. From this perspective,
the full collection of termwise contributions $\{a_{\alpha,k}\}$ provides the
microscopic representation on which our termwise gradient diagnostics are
built.

\subsection{Cancellation Ratio, Effective Term Count, and Sign Organization}

To quantify this cancellation structure, we introduce three complementary
diagnostics at the level of each parameter $\theta_k$. First, the
\emph{cancellation ratio} is defined as
 \begin{equation}\label{eq:Rk}
R_k = \frac{\left|\sum_\alpha a_{\alpha,k}\right|}{\sum_\alpha |a_{\alpha,k}|},
\end{equation}
which measures how much of the total signed contribution survives after
cancellation relative to the total unsigned activity. By construction, $0 \le
R_k \le 1$: values near $1$ indicate that most termwise contributions are
sign-aligned, whereas values near $0$ indicate strong cancellation.
Second, to quantify how broadly the gradient is distributed across Hamiltonian
terms, we define the \emph{effective term count}
\begin{equation}\label{eq:Neff}
N_{\mathrm{eff},k}
=
\frac{\left(\sum_\alpha |a_{\alpha,k}|\right)^2}{\sum_\alpha a_{\alpha,k}^2}.
\end{equation}
This quantity plays a role analogous to a participation-ratio measure: it is
small when only a few terms dominate the gradient and large when many terms
contribute with comparable magnitude. Finally, we combine these two aspects into
the \emph{organization diagnostic}
 \begin{equation}\label{eq:Beff}
B_{\mathrm{eff},k}=R_k\sqrt{N_{\mathrm{eff},k}}.
\end{equation}
The motivation for this definition is that $R_k$ alone quantifies cancellation
severity but does not distinguish why cancellation is weak. A large $R_k$ may
indicate structured sign alignment among many active Hamiltonian terms, but it may
also arise trivially when only a few terms participate and little cancellation
is possible. By combining $R_k$ with the participation breadth measured by
$N_{\mathrm{eff},k}$, the organization diagnostic
$B_{\mathrm{eff},k}=R_k\sqrt{N_{\mathrm{eff},k}}$ separates these cases. Thus,
$B_{\mathrm{eff},k}$ should be read not as a direct measure of gradient size,
but as a measure of structured sign organization beyond mere term-count
reduction.

\subsection{Conditional Random-Sign Model}

As a reference model for unstructured cancellation, we consider a
\emph{random-sign} description of the termwise gradient contributions. For a
fixed parameter $\theta_k$, we write
\begin{equation}
a_{\alpha,k}=s_{\alpha,k}w_{\alpha,k},
\end{equation}
where $w_{\alpha,k}\ge 0$ denotes the magnitude of the contribution from term
$\alpha$ for parameter $\theta_k$, and $s_{\alpha,k}\in\{+1,-1\}$ is a sign
variable. In the random-sign model, the signs $s_{\alpha,k}$ are treated as
independent symmetric random variables, so that positive and negative
contributions occur with no structural bias or correlation. The magnitudes
$\{w_{\alpha,k}\}$ are allowed to be nonuniform, and therefore the model does
not assume equal contribution strengths across Hamiltonian terms. Rather, it
isolates the hypothesis that cancellation is driven by essentially random sign
assignment on top of a given magnitude profile. This model provides a natural
null baseline for our framework: if an ansatz does not organize the signs of its
termwise contributions in any structured way, then its gradient should behave
approximately like a weighted random sum under this random-sign description.

From Eqs.~\eqref{eq:Rk}, \eqref{eq:Neff}, and \eqref{eq:Beff},
$B_{\mathrm{eff},k}$ can be written in the equivalent form
\begin{equation}\label{eq:Beff_equiv}
B_{\mathrm{eff},k}
=
\frac{\left|\sum_\alpha a_{\alpha,k}\right|}
     {\sqrt{\sum_\alpha a_{\alpha,k}^2}}.
\end{equation}
Under the decomposition $a_{\alpha,k}=s_{\alpha,k}\,w_{\alpha,k}$ with
$w_{\alpha,k}\ge 0$, this becomes
\begin{equation}
B_{\mathrm{eff},k}
=
\frac{\left|\sum_\alpha s_{\alpha,k}\,w_{\alpha,k}\right|}
     {\sqrt{\sum_\alpha w_{\alpha,k}^2}}.
\end{equation}
Conditioned on a fixed magnitude profile $\{w_{\alpha,k}\}$, the numerator is
the absolute value of a weighted random sum with zero mean and variance
$\sum_\alpha w_{\alpha,k}^2$.  For a sufficiently large number of contributing
terms, and provided that no single weight dominates the sum, the central limit
effect gives
\begin{equation}
\mathbb{E}\!\left[
  \left|\sum_\alpha s_{\alpha,k}\,w_{\alpha,k}\right|
  \;\middle|\;
  \{w_{\alpha,k}\}
\right]
\;\approx\;
\sqrt{\frac{2}{\pi}}\,\sqrt{\sum_\alpha w_{\alpha,k}^2}.
\end{equation}
Dividing both sides by $\sqrt{\sum_\alpha w_{\alpha,k}^2}$ then yields
\begin{equation}
\mathbb{E}\!\left[
  B_{\mathrm{eff},k}
  \;\middle|\;
  \{w_{\alpha,k}\}
\right]
\;\approx\;
\sqrt{\frac{2}{\pi}}.
\label{eq:beff_baseline}
\end{equation}
This baseline should be interpreted as an approximate finite-size reference
rather than an exact finite-sample law. In particular, when the effective number
of contributing terms is modest or the weight profile is strongly uneven,
deviations from the Gaussian approximation can be non-negligible. In the regimes
studied here, $N_{\mathrm{eff},k}$ is roughly in the single-digit to low-teen
range, so $\sqrt{2/\pi}$ should be read as a qualitative random-sign baseline
rather than an exact quantitative prediction.

Since the same conditional approximation is obtained for any fixed magnitude
profile, it follows that, under the random-sign model,
\begin{equation}
\mathbb{E}[B_{\mathrm{eff},k}]
\;\approx\;
\sqrt{\frac{2}{\pi}}.
\end{equation}
As a corollary, the expected cancellation ratio under the same conditioning satisfies
\begin{equation}
\mathbb{E}\!\left[
  R_k
  \;\middle|\;
  \{w_{\alpha,k}\}
\right]
\;\approx\;
\sqrt{\frac{2}{\pi}}\,
\frac{1}{\sqrt{N_{\mathrm{eff},k}}},
\label{eq:Rk_baseline}
\end{equation}
which provides the corresponding parameter-level random-sign
reference.

While the first-moment baseline above rests on a central limit
approximation, the second moment of $B_{\mathrm{eff},k}$ obeys an
exact identity under the same conditioning. Writing
$B_{\mathrm{eff},k}^2 = \left(\sum_\alpha s_{\alpha,k}
w_{\alpha,k}\right)^2 / \sum_\alpha w_{\alpha,k}^2$ and using the
independence and symmetry of the signs, the expectation of the
squared sum reduces to $\sum_\alpha w_{\alpha,k}^2$ because all
cross terms vanish in expectation. Hence
\begin{equation}
\mathbb{E}\!\left[
  B_{\mathrm{eff},k}^2
  \;\middle|\;
  \{w_{\alpha,k}\}
\right]
=
1,
\label{eq:B2_baseline}
\end{equation}
for every fixed magnitude profile with $\sum_\alpha w_{\alpha,k}^2 >
0$, with no appeal to the central limit theorem and no condition on
the number of terms or the balance of the weights. The two baselines
must not be conflated. The second moment satisfies
Eq.~\eqref{eq:B2_baseline} exactly, so Jensen's inequality gives
$\left(\mathbb{E}[B_{\mathrm{eff},k}]\right)^2 \le 1$ in general,
and the central limit approximation locates the first moment near
$\sqrt{2/\pi}$, whose square $2/\pi \approx 0.64$ lies well below
the exact second-moment value of $1$.
The exact second-moment value plays the central role in what follows,
since it is the level at which the random-sign model connects to the
gradient second moment through the variance bridge of
Section~\ref{s:var_bridge}, and it serves as the reference value
against which the measured organization factor is compared in
Section~\ref{s:decomposition}.

These baselines motivate using the random-sign model as a reference
for unstructured cancellation rather than as a literal microscopic
description of the gradient signs. At the aggregate level, the model
gives the approximate first-moment reference
$\mathbb{E}[B_{\mathrm{eff},k}] \approx \sqrt{2/\pi}$ under
CLT-like conditions and the exact second-moment reference of
Eq.~\eqref{eq:B2_baseline} when conditioned on a fixed magnitude
profile. Agreement with these references indicates that the net sign
organization is close to that expected from unstructured random
signs, whereas systematic departures indicate additional
organization. Such agreement does not imply that the microscopic
signs are themselves independent and symmetric; weak residual sign
structure may remain without accumulating into the aggregate
organization diagnostic. We use these references below to compare
the organization of different ansatz families across system sizes
and depths.

\section{Variance Bridge}
\label{s:var_bridge}

The diagnostics developed above quantify signed cancellation at the
level of termwise gradient contributions. We now connect them to
the standard variance-based description of BPs through an exact
identity for the gradient second moment, which resolves it into an
activity scale, an organization factor, and their coupling, and
which turns the conditional random-sign model into a source of
falsifiable predictions for two of the three factors.

\subsection{Exact Second-Moment Identity and Connection to Gradient Variance}

Let $a_{\alpha,k}$ denote the termwise gradient contributions introduced in
Eq.~\eqref{eq:termwise_grad}, and define
\begin{equation}
Q_k = \sum_\alpha a_{\alpha,k}^2.
\end{equation}
Here, $Q_k$ quantifies the pre-cancellation activity of the termwise
gradient contributions. Using Eq.~\eqref{eq:Beff_equiv}, we obtain
\begin{equation}
\left(\sum_\alpha a_{\alpha,k}\right)^2
=
B_{\mathrm{eff},k}^2 Q_k.
\label{eq:variance_bridge_identity}
\end{equation}
This pointwise identity is exact for every parameter and every run.
It expresses the squared aggregate gradient as the product of the
activity scale $Q_k$ and the squared organization diagnostic
$B_{\mathrm{eff},k}^2$, without assuming which quantity carries the
dependence on system size.

Taking the expectation over parameter initialization or circuit
ensemble gives the exact second-moment relation
\begin{equation}
\mathbb{E}_{\theta}\!\left[(\partial_k\langle H\rangle)^2\right]
=
\mathbb{E}_{\theta}\!\left[B_{\mathrm{eff},k}^2 Q_k\right].
\end{equation}
The corresponding gradient variance is therefore
\begin{equation}
\mathrm{Var}_{\theta}[\partial_k\langle H\rangle]
=
\mathbb{E}_{\theta}\!\left[B_{\mathrm{eff},k}^2 Q_k\right]
-
\left(
\mathbb{E}_{\theta}[\partial_k\langle H\rangle]
\right)^2.
\label{eq:variance_bridge_variance}
\end{equation}
Thus, the connection from the exact second-moment identity to the
variance requires only the explicit mean-gradient correction in the
second term. We quantify this correction rather than assume it to
vanish: its group-level counterpart, which enters the variance object of 
Eq.~\eqref{eq:var_def}, is estimated with finite-sample bias
correction at the end of Section~\ref{s:decomposition} and is
consistent with zero in every tested $(n,d)$ condition for both
ansatz families and both Hamiltonians.

Equation~\eqref{eq:variance_bridge_variance}, together with the
exact second-moment relation above, constitutes the variance bridge
of this work: the exact second-moment term is resolved into
pre-cancellation activity and sign organization through
$B_{\mathrm{eff},k}^2 Q_k$, while the remaining mean-gradient term
specifies the correction required to obtain the variance.
This relation does not by itself identify the source of gradient
suppression. Rather, it makes explicit that changes in the gradient
second moment can arise from changes in termwise activity, changes
in sign organization, or their statistical coupling. 

The algebraic identity itself is not the empirical result of this
work. Its role is to provide a lossless starting point from which
the gradient second moment can be analyzed in terms of the
group-level activity factor $\mathbb{E}[Q_k]$, the organization
factor $\mathbb{E}[B_{\mathrm{eff},k}^2]$, and their coupling $F$.
The empirical content lies in determining which of these factors
carries the dependence on system size and in testing the reference
predictions supplied by the conditional random-sign model, namely
$\mathbb{E}[B_{\mathrm{eff},k}^2]=1$ and $F=1$. Both predictions
could have failed in the data, so their status is an empirical
question rather than a consequence of the decomposition itself.
The corresponding estimands and estimation procedures are specified
in Sections~\ref{ss:termwise_methods} and~\ref{ss:statistics}.

\subsection{Organization--Activity Coupling under the Random-Sign Model}
\label{ss:factorization}

The exact second-moment relation resolves
$\mathbb{E}_{\theta}[B_{\mathrm{eff},k}^2 Q_k]$ no further. A
natural question is when this product moment factorizes,
\begin{equation}
\mathbb{E}_{\theta}[B_{\mathrm{eff},k}^2 Q_k]
\approx
\mathbb{E}_{\theta}[B_{\mathrm{eff},k}^2]\,
\mathbb{E}_{\theta}[Q_k].
\label{eq:factorization_bridge}
\end{equation}
To quantify the accuracy of this factorization, we denote by
$F = \mathbb{E}_{\theta}[B_{\mathrm{eff},k}^2 Q_k] \big/
\big(\mathbb{E}_{\theta}[B_{\mathrm{eff},k}^2]\,
\mathbb{E}_{\theta}[Q_k]\big)$
the ratio of the exact product moment to its factorized
counterpart, so that $F = 1$ corresponds to exact factorization.
Its treatment as a formal estimand, on the same footing as the other
decomposition factors, is given in Sec.~\ref{ss:termwise_methods}.
This factorization is not exact in general. Under the conditional
random-sign model of Section~\ref{s:framework}, however, $F = 1$ is
an exact prediction rather than an approximation. Conditioned on a
magnitude profile $\{w_{\alpha,k}\}$, the activity
$Q_k = \sum_\alpha w_{\alpha,k}^2$ is fixed, so
Eq.~\eqref{eq:B2_baseline} gives
$\mathbb{E}\big[B_{\mathrm{eff},k}^2 Q_k \,\big|\,
\{w_{\alpha,k}\}\big] = Q_k$. Averaging over profiles then yields
$\mathbb{E}[B_{\mathrm{eff},k}^2 Q_k] = \mathbb{E}[Q_k]$ together
with $\mathbb{E}[B_{\mathrm{eff},k}^2] = 1$, and hence $F = 1$
exactly. Departures of $F$ from 1 are therefore a falsifiable
signature of coupling between sign organization and activity beyond
the random-sign description.
Empirically, the coupling factors measured in Section~\ref{s:decomposition} are
consistent with this prediction for the HEA in both Hamiltonians, whereas the
HVA shows departures from $F = 1$ toward larger systems,
quantitatively stronger in the LFIM.

The conceptual consequence is important. 
The three-factor decomposition characterizes the gradient second moment
into the group-level activity factor $\mathbb{E}[Q_k]$, the
organization factor $\mathbb{E}[B_{\mathrm{eff},k}^2]$, and their
coupling $F$, and ansatz families can differ in any of the three.

\subsection{Relation to Lie-Algebraic Approaches}
\label{ss:lie}

Recent work has analyzed BPs through the lens of Lie algebra theory, relating
the dynamical Lie algebra (DLA) generated by a parameterized circuit to
gradient-variance scaling under suitable expressibility, mixing, and depth
conditions~\cite{fontana:2024:natcomm,ragone:2024:natcomm}. These approaches
provide a structural account of \emph{why} certain circuit families become
susceptible to variance suppression: as the generated algebraic directions
spread and mix sufficiently, gradients concentrate in a manner characteristic of
BP behavior. Our framework addresses a different but complementary question.
Rather than characterizing the algebraic reach of the circuit, we ask how the
resulting gradient signal is organized across the Hamiltonian terms that define
the objective. In this sense, Lie-algebraic analyses provide a circuit-level
structural account of variance suppression, whereas our diagnostics
provide a term-resolved account of how the gradient signal is
composed and organized across the signed Hamiltonian-term contributions. 

This distinction also clarifies the role of the diagnostics
introduced here. They are not intended to replace DLA-based
structural criteria, and we do not claim that they proxy for DLA
dimension. Instead, they operate inside the variance picture: the
decomposition resolves the gradient second moment into the activity
scale, the organization factor, and their coupling. Two circuits
may both exhibit substantial algebraic spreading and yet differ in
how their termwise contributions are organized in sign, and this
difference is made measurable by the organization factor and by the
sign-alignment analysis of Section~\ref{s:microscopic}.

This perspective is particularly relevant for structured ansatzes.
Lie-algebraic and locality-based analyses suggest that
problem-informed ansatzes can avoid or delay BPs when their
evolution remains sufficiently tied to local or
Hamiltonian-generated structure. For HVAs in particular,
Park~\cite{park:2024:quantum} shows that BPs can be avoided when
the circuit is well approximated by local-Hamiltonian time
evolution. Our framework contributes a termwise reading of the same
setting: if the ansatz generators remain aligned with the operator
sectors of the Hamiltonian, their termwise contributions need not
behave as unstructured random-sign variables, and whether they do
is a measurable question. The sector-structured sign alignment
documented in Section~\ref{s:microscopic} is consistent with this
reading, within the comparison-design limits discussed there. The
relevant issue, in our interpretation, is then not DLA dimension
alone but the compatibility between the generated directions and
the Hamiltonian sectors. Establishing a formal connection between
DLA structure, generator--Hamiltonian alignment, and the
organization factor remains an important direction for future work.

\section{Decomposition of Gradient Second-Moment Scaling}
\label{s:decomposition}

The preceding sections established that the gradient second moment
decomposes exactly into an organization factor, an activity factor,
and their coupling, and that the conditional random-sign model makes
falsifiable predictions for two of the three. This section carries
out the corresponding empirical program. 
We first establish the contrast in gradient-variance scaling
between the two ansatz families, with the HEA settings exhibiting
the standard suppression symptom of BP behavior, 
then summarize the diagnostic profile of the two families, and
finally decompose the scaling of the gradient second moment $M_2$,
the squared gradient averaged over parameters and initializations
and defined formally in Eq.~\eqref{eq:M2_def}, to determine which
factors carry it.

\subsection{Ansatz Families and Comparison Design}

We compare two circuit families on the TFIM with open boundary conditions,
\begin{equation}
H
=
-\sum_{i=1}^{n-1} Z_i Z_{i+1}
+
h\sum_{i=1}^{n} X_i,
\end{equation}
with $h = 1$, and in parallel on the LFIM, the longitudinal-field 
extension defined in Eq.~\eqref{eq:lfim_hamiltonian}. The HEA is a
generic expressive circuit with weak problem alignment, and the HVA
is a problem-informed circuit whose layer generators are drawn from
the operator sectors of the target Hamiltonian. Full circuit
constructions, parameter counts, and the experimental grid are given
in Sec.~\ref{ss:circuit_definitions}.

This pairing makes the comparison informative. Both families are
evaluated on the same cost function over the same grid of system
sizes and depths, so the differences in their gradient statistics
cannot be attributed to the objective and instead reflect
differences between the circuit families. These families differ in
several respects, including parameter count and parameter sharing,
and their principal structural contrast is whether the layer
generators remain aligned with the Hamiltonian sectors. Throughout
this section the TFIM serves as the primary case and the LFIM as a
parallel robustness test, with both Hamiltonians analyzed by the
identical pipeline.

\subsection{Gradient-Variance Scaling Check}

Before interpreting the termwise diagnostics, we first examine the
gradient-variance scaling of the two ansatz families in the standard
BP sense. For each parameter $\theta_k$, we computed the variance of
$\partial_k\langle H\rangle$ across random seeds and then averaged
this quantity over parameters within each $(n,d,\mathrm{variant})$
condition. This estimates the variance object
$\mathbb{E}_k[\mathrm{Var}_s(\partial_k\langle H\rangle)]$ of
Eq.~\eqref{eq:var_def}, so the quantity anchoring this section is
the same one whose relation to $M_2$ is quantified by the
mean-gradient correction at the end of the section.

\begin{figure}[t]
    \centering
    \includegraphics[width=1.0\textwidth]{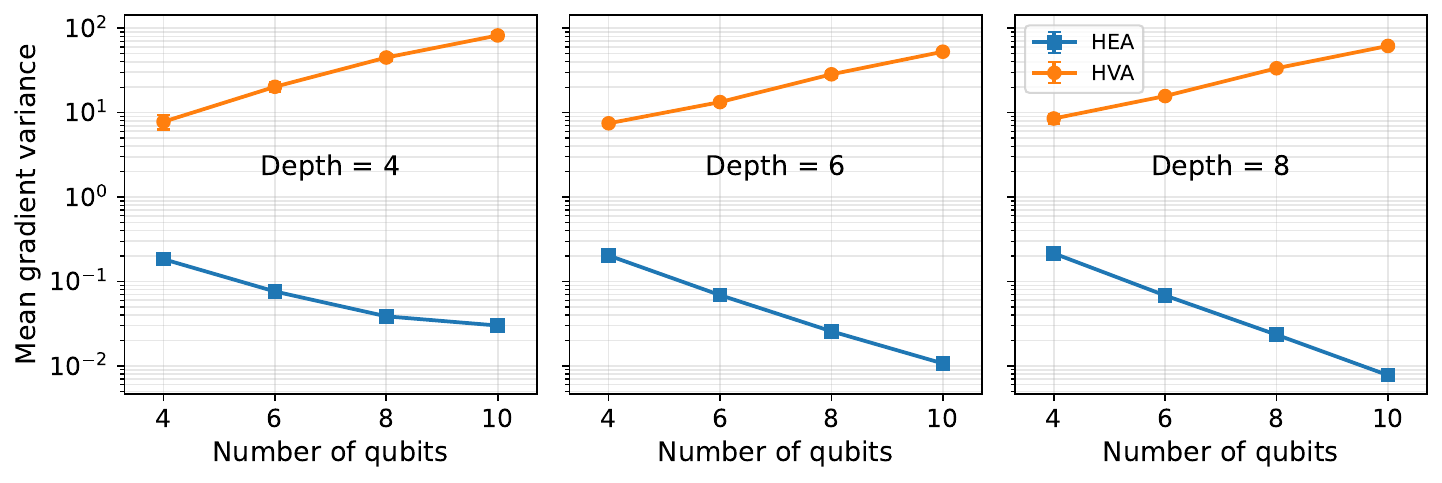}
    \caption{Finite-size gradient-variance scaling for HEA and HVA across circuit depths in the TFIM. The plotted quantity is the mean parameter-wise variance of $\partial_k\langle H\rangle$, computed across random seeds and then averaged over parameters for each $(n,d,\mathrm{variant})$ condition. HEA shows a clear decrease in gradient variance as the number of qubits increases, whereas HVA instead shows increasing variance over the tested range. This provides a standard variance-based check that the HEA settings exhibit BP-relevant gradient suppression, while HVA retains substantially larger gradient signal.}
    \label{fig:grad_variance}
\end{figure}

Figure~\ref{fig:grad_variance} shows a clear contrast between the two ansatz
families. For HEA, the mean gradient variance decreases monotonically with
system size at all tested depths, with the suppression becoming stronger at
larger circuit depth. By contrast, HVA does not show such suppression over the
same range. Instead, its gradient variance increases with $n$. This behavior is
consistent with the interpretation that HEA enters a BP-relevant finite-size
gradient-suppression regime, whereas HVA preserves substantially larger gradient
variance under the tested conditions.

This variance-based check complements the decomposition that
follows. The goal is not to claim an asymptotic BP scaling law from
the finite range $n\in\{4,6,8,10\}$, but to establish the
finite-size variance-scaling contrast whose corresponding
second-moment scaling the remainder of this section decomposes: the
diagnostic profile summarized next provides continuity with the
framework of Section~\ref{s:framework}, the second-moment
decomposition then determines which factors carry the scaling, and
the closing mean-gradient check quantifies how closely the
second-moment conclusions transfer to the variance.

\subsection{Diagnostic Profile of the Two Ansatz Families}

At the diagnostic level, the two families separate cleanly. 
HEA remains within a relatively stable low-$B_{\mathrm{eff}}$ band 
across the tested system sizes and depths (Tables~\ref{tab:hea_hva_summary}
and~\ref{tab:lfim_hea_hva_summary}), without the systematic upward 
shift that would indicate a departure from random-sign behavior. 

HVA departs from this band systematically. Across all tested
$(n,d)$ conditions it attains larger $B_{\mathrm{eff}}$ than HEA,
the bootstrap 95\% confidence interval for
$\Delta B_{\mathrm{eff}} = B_{\mathrm{eff}}^{\mathrm{HVA}} -
B_{\mathrm{eff}}^{\mathrm{HEA}}$ excludes zero in every condition,
and the gap widens with system size at each tested depth.
This difference in $B_{\mathrm{eff}}$ is not a term-count artifact.
A large $R_k$ could arise trivially if few
terms were active, but HVA exhibits larger $N_{\mathrm{eff}}$
together with larger $R_k$ and $B_{\mathrm{eff}}$
(Table~\ref{tab:hea_hva_summary}), so broader participation and
stronger sign organization occur simultaneously, which rules out the
sparsification explanation.

\setlength{\tabcolsep}{11pt}
\begin{table}[t]
\centering
\caption{Parameter-level diagnostic comparison of HEA and HVA for the TFIM across system size ($n$) and circuit depth ($d$). The table reports the cancellation ratio $R$, effective term count $N_{\mathrm{eff}}$, and organization diagnostic $B_{\mathrm{eff}}$. HVA exhibits larger $B_{\mathrm{eff}}$ while also involving a larger effective number of contributing Hamiltonian terms, showing that its stronger sign organization is not explained by trivial sparsification.}
\label{tab:hea_hva_summary}
\begin{tabular}{ccccccccc}
\toprule
$n$ & $d$ & $R_{\mathrm{HEA}}$ & $R_{\mathrm{HVA}}$ & $N_{\mathrm{eff}}^{\mathrm{HEA}}$ & $N_{\mathrm{eff}}^{\mathrm{HVA}}$ & $B_{\mathrm{eff}}^{\mathrm{HEA}}$ & $B_{\mathrm{eff}}^{\mathrm{HVA}}$ & $\Delta B_{\mathrm{eff}}$ \\
\midrule
4  & 4 & 0.462 & 0.505 & 3.908 & 4.811  & 0.853 & 1.069 & 0.215 \\
4  & 6 & 0.441 & 0.534 & 4.283 & 4.753  & 0.861 & 1.123 & 0.262 \\
4  & 8 & 0.435 & 0.543 & 4.394 & 4.859  & 0.870 & 1.166 & 0.295 \\
6  & 4 & 0.445 & 0.513 & 4.899 & 7.493  & 0.881 & 1.370 & 0.489 \\
6  & 6 & 0.378 & 0.444 & 5.944 & 7.541  & 0.835 & 1.184 & 0.349 \\
6  & 8 & 0.349 & 0.478 & 6.357 & 7.386  & 0.808 & 1.273 & 0.465 \\
8  & 4 & 0.422 & 0.551 & 5.570 & 10.229 & 0.850 & 1.722 & 0.872 \\
8  & 6 & 0.358 & 0.464 & 7.447 & 10.330 & 0.852 & 1.460 & 0.608 \\
8  & 8 & 0.321 & 0.496 & 8.184 & 10.174 & 0.816 & 1.548 & 0.733 \\
10 & 4 & 0.427 & 0.582 & 5.523 & 13.038 & 0.844 & 2.055 & 1.211 \\
10 & 6 & 0.340 & 0.495 & 8.545 & 13.166 & 0.828 & 1.762 & 0.934 \\
10 & 8 & 0.303 & 0.527 & 9.946 & 13.027 & 0.823 & 1.866 & 1.043 \\
\bottomrule
\end{tabular}
\end{table}

The same diagnostic pattern holds in the LFIM
(Table~\ref{tab:lfim_hea_hva_summary}): HVA exhibits larger $R$,
larger $N_{\mathrm{eff}}$, and larger $B_{\mathrm{eff}}$ in all
twelve conditions, broadening the evidence beyond a single
Hamiltonian. These observations establish that the diagnostic
profiles of the two families differ, but they do not yet determine
which factors carry the scaling of the gradient second moment with
system size. That question is answered by the decomposition that
follows.

\setlength{\tabcolsep}{11pt}
\begin{table}[!htbp]
\centering
\caption{Parameter-level diagnostic comparison of HEA and HVA for the LFIM across system size ($n$) and circuit depth ($d$). The table reports the cancellation ratio $R$, effective term count $N_{\mathrm{eff}}$, and organization diagnostic $B_{\mathrm{eff}}$. As in the TFIM, HVA exhibits larger $B_{\mathrm{eff}}$ while also involving a larger effective number of contributing Hamiltonian terms, showing that the stronger organization diagnostic is not explained by trivial sparsification.}
\label{tab:lfim_hea_hva_summary}
\begin{tabular}{ccccccccc}
\toprule
$n$ & $d$ & $R_{\mathrm{HEA}}$ & $R_{\mathrm{HVA}}$ & $N_{\mathrm{eff}}^{\mathrm{HEA}}$ & $N_{\mathrm{eff}}^{\mathrm{HVA}}$ & $B_{\mathrm{eff}}^{\mathrm{HEA}}$ & $B_{\mathrm{eff}}^{\mathrm{HVA}}$ & $\Delta B_{\mathrm{eff}}$ \\
\midrule
4  & 4 & 0.381 & 0.455 & 5.581 & 6.660  & 0.826 & 1.112 & 0.286 \\
4  & 6 & 0.362 & 0.430 & 6.126 & 6.827  & 0.832 & 1.072 & 0.241 \\
4  & 8 & 0.354 & 0.469 & 6.318 & 6.778  & 0.839 & 1.172 & 0.333 \\
6  & 4 & 0.375 & 0.434 & 7.168 & 10.217 & 0.877 & 1.323 & 0.446 \\
6  & 6 & 0.316 & 0.412 & 8.630 & 10.373 & 0.833 & 1.275 & 0.442 \\
6  & 8 & 0.296 & 0.432 & 9.154 & 10.386 & 0.818 & 1.343 & 0.525 \\
8  & 4 & 0.353 & 0.468 & 8.109 & 13.898 & 0.840 & 1.663 & 0.823 \\
8  & 6 & 0.295 & 0.430 & 10.897 & 14.111 & 0.845 & 1.559 & 0.713 \\
8  & 8 & 0.264 & 0.444 & 11.866 & 14.058 & 0.812 & 1.614 & 0.802 \\
10 & 4 & 0.359 & 0.495 & 7.883 & 17.626 & 0.827 & 1.976 & 1.149 \\
10 & 6 & 0.281 & 0.456 & 12.300 & 17.921 & 0.822 & 1.866 & 1.044 \\
10 & 8 & 0.250 & 0.462 & 14.369 & 17.811 & 0.818 & 1.893 & 1.075 \\
\bottomrule
\end{tabular}
\end{table}

\subsection{Scaling of Activity and Organization}

\begin{figure}[!htbp]
    \centering
    \includegraphics[width=1.0\textwidth]{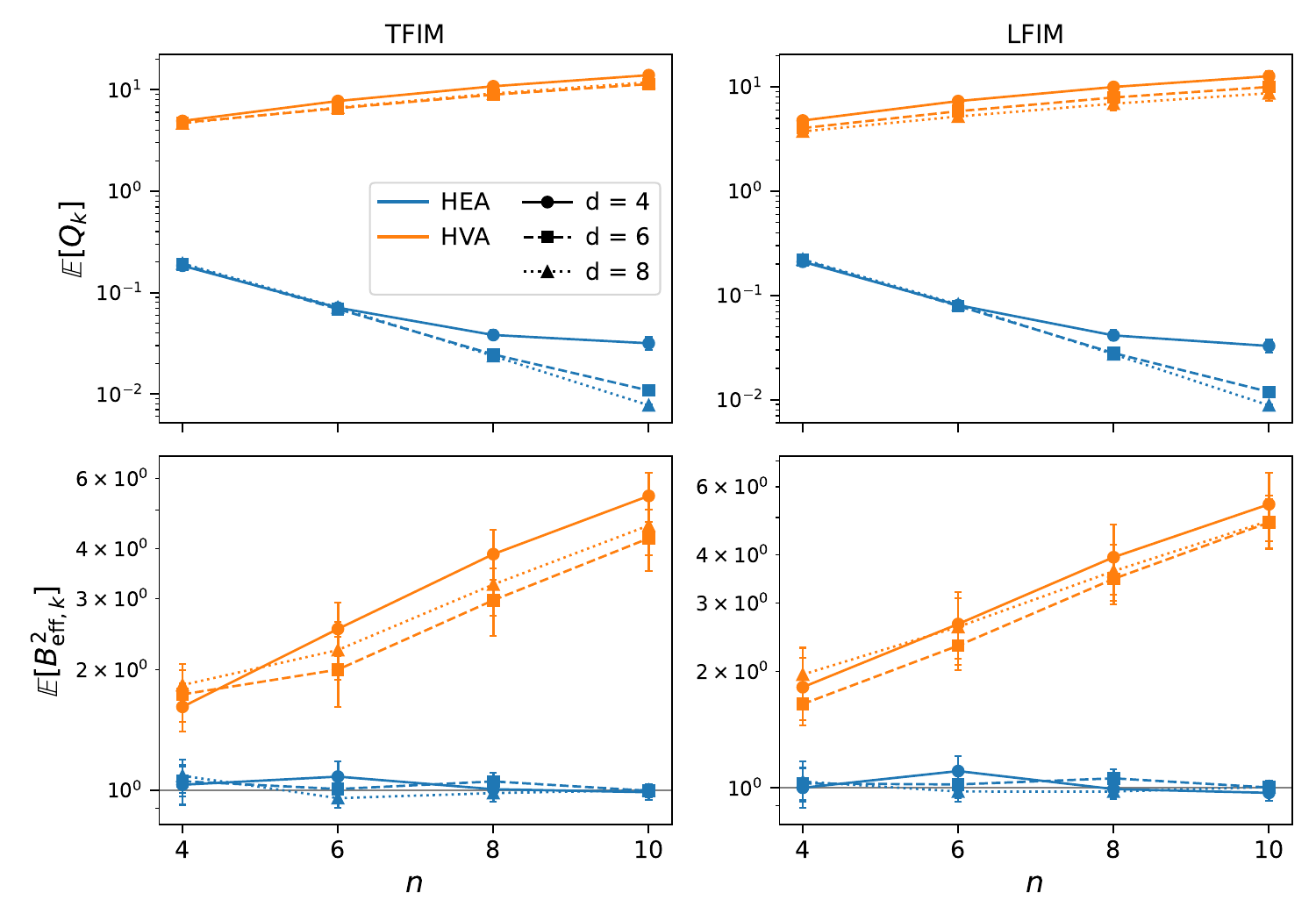}
    \caption{Scaling of the activity and organization factors with
    system size for the TFIM (left) and the LFIM (right). Top:
    activity scale $\mathbb{E}[Q_k]$. Bottom: organization factor
    $\mathbb{E}[B_{\mathrm{eff},k}^2]$, with the gray line marking
    the exact random-sign prediction
    $\mathbb{E}[B_{\mathrm{eff},k}^2] = 1$ of
    Eq.~\eqref{eq:B2_baseline}. Error bars are bootstrap 95\%
    confidence intervals over seeds, and the vertical axes are
    logarithmic. In both Hamiltonians the HEA activity decreases with
    system size while its organization factor stays consistent with
    the random-sign value, whereas for the HVA both factors grow.}
    \label{fig:q_b2_scaling}
\end{figure}

We now decompose the second-moment scaling into its factors,
beginning with the activity scale $\mathbb{E}[Q_k]$ and the
organization factor $\mathbb{E}[B_{\mathrm{eff},k}^2]$ separately.
Figure~\ref{fig:q_b2_scaling} shows both quantities as functions of
system size for each depth and ansatz family, for the TFIM and the
LFIM in parallel. 
Throughout, scaling is summarized by log-linear
fits over the tested range $n \in \{4,\dots,10\}$, with one
deviation noted below at the shallowest depth, and we make no claim
about asymptotic behavior.

The activity scale separates the two families sharply. For HEA, the
per-qubit decay factor of $\mathbb{E}[Q_k]$ is $0.744$, $0.618$, and
$0.584$ at depths $4$, $6$, and $8$ in the TFIM, with bootstrap
slope confidence intervals excluding zero, and the LFIM values are
nearly identical ($0.732$, $0.613$, $0.584$). 
At the shallowest depth the decay flattens toward larger $n$ in both
Hamiltonians, consistent with a shallow-depth effect reproduced
across the two tested models, although this alone does not establish
that the effect is Hamiltonian-independent.
For HVA, by contrast,
$\mathbb{E}[Q_k]$ grows with system size, with per-qubit factors of
about $1.16$ to $1.19$ in the TFIM and $1.15$ to $1.18$ in the
LFIM. 
Termwise gradient activity therefore decreases with system size only
in the hardware-efficient family over the tested range.

The organization factor shows the complementary contrast. For HEA,
$\mathbb{E}[B_{\mathrm{eff},k}^2]$ remains flat and consistent, 
at the second-moment level, with the exact random-sign prediction 
of Eq.~\eqref{eq:B2_baseline}: the
group means lie in $[0.955, 1.087]$ across all twelve TFIM
conditions and in $[0.971, 1.105]$ in the LFIM, and the slope
confidence intervals all include zero. For HVA,
$\mathbb{E}[B_{\mathrm{eff},k}^2]$ instead grows with system size,
from about $1.6$ to values between $4.2$ and $5.4$ across depths in
the TFIM, with per-qubit factors of $1.17$ to $1.23$, and with
closely similar growth in the LFIM ($1.17$ to $1.20$). 
This growth occurs while participation broadens and sign alignment
remains substantial: the mean squared cancellation ratio $\overline{R_k^2}$
stays of order one ($0.28$ to $0.43$ in the TFIM, $0.24$ to $0.34$
in the LFIM) while $N_{\mathrm{eff}}$ grows from about $5$ to $13$
(TFIM) and from about $7$ to $18$ (LFIM), so
$B_{\mathrm{eff},k}^2 = R_k^2 N_{\mathrm{eff},k}$ grows without
requiring any systematic increase in $R_k^2$ itself.

Taken together, the two factors already indicate qualitatively
different scaling compositions: in HEA the activity decreases while
the organization sits at the random-sign level, whereas in HVA both
factors grow. The quantitative apportioning of the second-moment
slope between them is the subject of the next subsection.

\subsection{Slope-Share Decomposition of the Second Moment}

\begin{figure}[!htbp]
    \centering
    \includegraphics[width=1.0\textwidth]{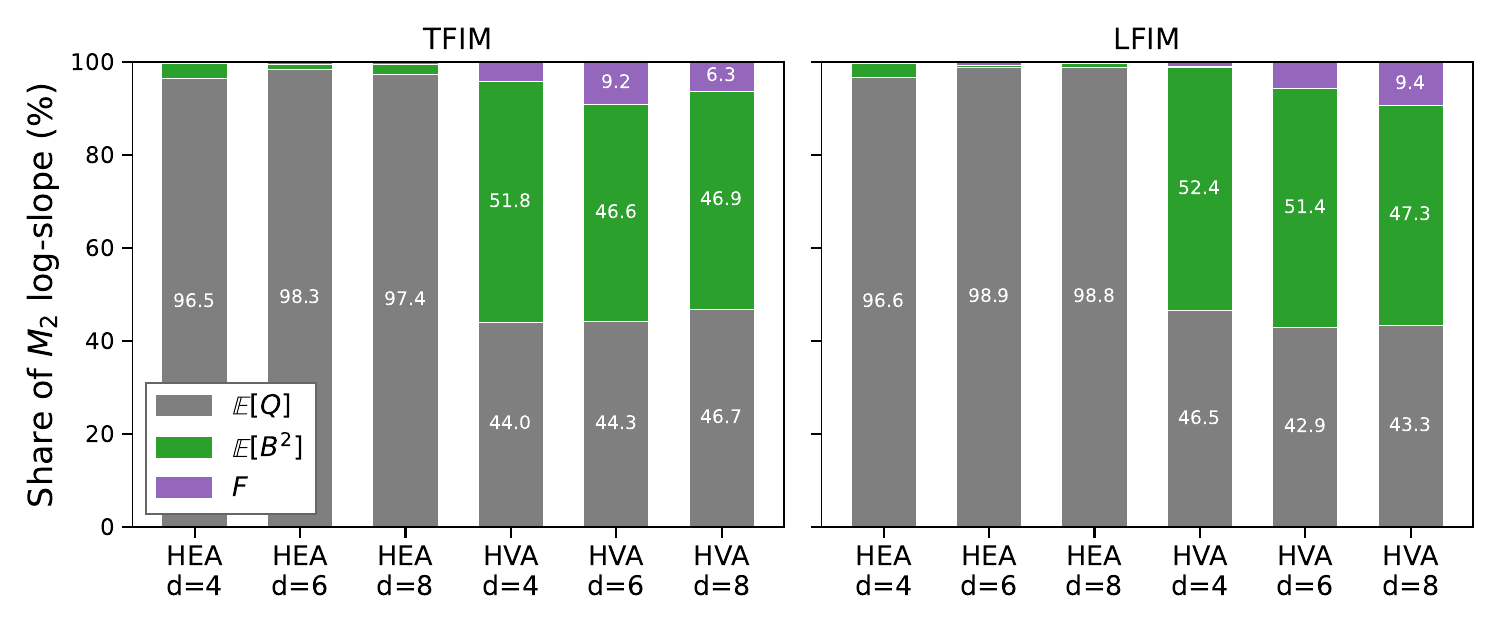}
    \caption{Slope-share decomposition of the gradient second moment
    for the TFIM (left) and the LFIM (right). Each bar decomposes the
    fitted $M_2$ log-slope of one ansatz family and depth into the
    contributions of the activity, organization, and coupling
    components, following Eq.~\eqref{eq:log_decomposition}. The
    shares add to $100\%$ exactly before rounding and describe the
    composition of the second-moment log change over the tested
    range, with no causal attribution. The HEA slope is carried
    almost entirely by the activity, whereas the HVA slope receives
    comparable contributions from activity and organization.}
    \label{fig:m2_decomposition}
\end{figure}

The additive form of Eq.~\eqref{eq:log_decomposition} makes the
qualitative separation of the previous subsection quantitative. For
each depth and ansatz family, we decompose the fitted log-slope of
$M_2$ with respect to system size into the slopes of
$\mathbb{E}[B_{\mathrm{eff},k}^2]$, $\mathbb{E}[Q_k]$, and $F$, and
report each component as a share of the total. The additivity is
exact on the data, since it follows from the identity of
Eq.~\eqref{eq:three_factor} applied to the fitted slopes.
Figure~\ref{fig:m2_decomposition} shows the decomposition for the
TFIM and the LFIM side by side, and
Table~\ref{tab:decomposition_summary} collects the per-qubit
factors, slope shares, and coupling factors for all conditions.

\begin{table}[!htbp]
\centering
\small
\caption{Decomposition of the gradient second-moment scaling for both
Hamiltonians. For each depth $d$ and ansatz family, the per-qubit
multiplicative factors ($e^{\mathrm{slope}}$ of the log-linear fit
over $n \in \{4,\dots,10\}$) are given for the activity
$\mathbb{E}[Q_k]$, the organization
$\mathbb{E}[B_{\mathrm{eff},k}^2]$, the coupling $F$, and their
product $M_2$, together with the share of each component slope in
the $M_2$ log-slope. By the identity of
Eq.~\eqref{eq:log_decomposition}, the factor columns multiply to the
$M_2$ column and the share columns add to $100\%$ exactly before
rounding, with displayed values subject to rounding at the last
digit.}
\label{tab:decomposition_summary}
\setlength{\tabcolsep}{9pt}
\begin{tabular}{llcccccccc}
\toprule
& & & \multicolumn{4}{c}{Per-qubit factor}
& \multicolumn{3}{c}{Slope share (\%)} \\
\cmidrule(lr){4-7}\cmidrule(lr){8-10}
Hamiltonian & Ansatz & $d$
& $\mathbb{E}[Q]$ & $\mathbb{E}[B^2]$ & $F$ & $M_2$
& $Q$ & $B^2$ & $F$ \\
\midrule
TFIM & HEA & 4 & 0.744 & 0.990 & 0.999 & 0.736 & 96.5 & 3.3 & 0.2 \\
TFIM & HEA & 6 & 0.618 & 0.994 & 0.998 & 0.613 & 98.3 & 1.2 & 0.4 \\
TFIM & HEA & 8 & 0.584 & 0.989 & 0.997 & 0.575 & 97.4 & 2.0 & 0.6 \\
TFIM & HVA & 4 & 1.188 & 1.225 & 1.017 & 1.480 & 44.0 & 51.8 & 4.2 \\
TFIM & HVA & 6 & 1.158 & 1.167 & 1.031 & 1.393 & 44.3 & 46.6 & 9.2 \\
TFIM & HVA & 8 & 1.168 & 1.169 & 1.021 & 1.395 & 46.7 & 46.9 & 6.3 \\
\midrule
LFIM & HEA & 4 & 0.732 & 0.990 & 0.999 & 0.724 & 96.6 & 3.1 & 0.4 \\
LFIM & HEA & 6 & 0.613 & 0.998 & 0.996 & 0.610 & 98.9 & 0.3 & 0.8 \\
LFIM & HEA & 8 & 0.584 & 0.995 & 0.998 & 0.580 & 98.8 & 0.9 & 0.3 \\
LFIM & HVA & 4 & 1.176 & 1.201 & 1.004 & 1.418 & 46.5 & 52.4 & 1.1 \\
LFIM & HVA & 6 & 1.164 & 1.200 & 1.020 & 1.424 & 42.9 & 51.4 & 5.7 \\
LFIM & HVA & 8 & 1.151 & 1.166 & 1.031 & 1.385 & 43.3 & 47.3 & 9.4 \\
\bottomrule
\end{tabular}
\end{table}

For HEA, the outcome is unambiguous. The activity slope accounts for
$96.5\%$, $98.3\%$, and $97.4\%$ of the $M_2$ log-slope at depths
$4$, $6$, and $8$ in the TFIM, and for $96.6\%$, $98.9\%$, and
$98.8\%$ in the LFIM. The organization and coupling components are
flat, with slope confidence intervals including zero, and $F$ itself
stays within $[0.966, 1.028]$ across all TFIM conditions. 
The $n$-dependence of the gradient second moment in HEA is therefore
carried almost entirely by the decay of termwise activity, while the
organization factor and the coupling factor stay near, and
statistically consistent with, their aggregate random-sign reference
values. At this aggregate diagnostic level, both falsifiable
predictions of the conditional random-sign model,
$\mathbb{E}[B_{\mathrm{eff},k}^2] = 1$ and $F = 1$, are consistent
with the data. A finer microscopic examination of the residual sign
structure is deferred to Section~\ref{s:microscopic}.

For HVA, no single factor dominates. In the TFIM, the activity and
organization shares are $44.0\%$ and $51.8\%$ at depth $4$,
$44.3\%$ and $46.6\%$ at depth $6$, and $46.7\%$ and $46.9\%$ at
depth $8$, with the coupling factor contributing the remaining
$4.2\%$, $9.2\%$, and $6.3\%$. The LFIM shares are closely parallel
($46.5/52.4/1.1$, $42.9/51.4/5.7$, and $43.3/47.3/9.4$ percent for
activity, organization, and coupling at the three depths).
Activity and organization therefore make comparable contributions to
the second-moment growth, with neither factor dominating, and the
coupling component is small but not negligible, reaching up to about
$9\%$ depending on depth and Hamiltonian.

Two reading rules apply to these numbers. The shares describe the
composition of the second-moment log change over the tested range
and carry no causal attribution. 
And the coupling component, while minor in the share accounting, is
where the clearest quantitative difference between the two
Hamiltonians appears, which is the subject of the next subsection.

\subsection{Organization--Activity Coupling}

\begin{figure}[!htbp]
    \centering
    \includegraphics[width=1.0\textwidth]{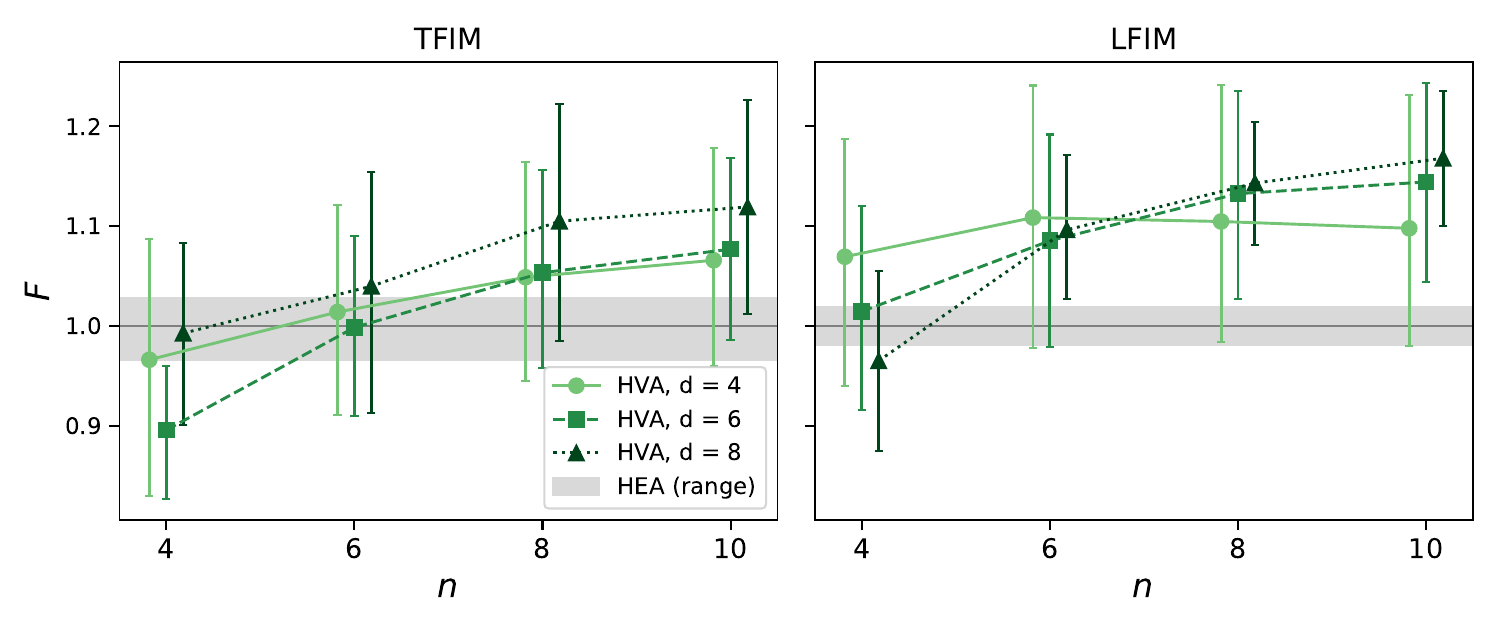}
    \caption{Coupling factor $F$ as a function of system size for
    the HVA at each depth, for the TFIM (left) and the LFIM (right).
    Error bars are bootstrap 95\% confidence intervals over seeds,
    and symbols are horizontally offset for readability. The gray
    line marks the random-sign prediction $F = 1$, and the shaded
    band spans the range of HEA values, all of whose confidence
    intervals contain $1$. The departures of $F$ from $1$ toward
    larger systems are quantitatively stronger in the LFIM.}
    \label{fig:coupling}
\end{figure}

The remaining factor is the coupling. By the identity of
Eq.~\eqref{eq:F_cov}, $F - 1$ equals the normalized covariance
between $B_{\mathrm{eff},k}^2$ and $Q_k$, and the law of total
covariance splits it exactly into a within-seed and a between-seed
component. Figure~\ref{fig:coupling} shows $F$ as a function of
system size with bootstrap confidence intervals, and 
Table~\ref{tab:coupling_channels} reports the two covariance 
components, for both Hamiltonians.

\begin{table}[!htbp]
\centering
\small
\caption{Within-seed and between-seed components of the normalized
$B_{\mathrm{eff}}^2$--$Q$ covariance for the HVA, for both
Hamiltonians. The two components add exactly to $F - 1$ before
rounding, by the plug-in law-of-total-covariance decomposition of
Sec.~\ref{ss:statistics}. Brackets give the bootstrap 95\%
confidence intervals over seeds, and bold entries mark components
whose interval excludes zero.}
\label{tab:coupling_channels}
\setlength{\tabcolsep}{10pt}
\begin{tabular}{cccrrr}
\toprule
Hamiltonian & $d$ & $n$ & $F-1$\ \ & \multicolumn{1}{c}{Within-seed} & \multicolumn{1}{c}{Between-seed} \\
\midrule
TFIM & 4 & 4  & $-0.034$ & $-0.079$ [$-0.189$, $0.030$] & $0.045$ [$0.000$, $0.090$] \\
TFIM & 4 & 6  & $+0.014$ & $-0.010$ [$-0.090$, $0.072$] & $0.024$ [$-0.034$, $0.082$] \\
TFIM & 4 & 8  & $+0.049$ & $0.023$ [$-0.068$, $0.115$] & $0.026$ [$-0.017$, $0.083$] \\
TFIM & 4 & 10 & $+0.066$ & $0.038$ [$-0.048$, $0.133$] & $0.028$ [$-0.014$, $0.078$] \\
TFIM & 6 & 4  & $-0.104$ & $\mathbf{-0.096}$ [$-0.148$, $-0.044$] & $-0.009$ [$-0.047$, $0.029$] \\
TFIM & 6 & 6  & $-0.002$ & $-0.014$ [$-0.085$, $0.057$] & $0.013$ [$-0.043$, $0.068$] \\
TFIM & 6 & 8  & $+0.053$ & $0.033$ [$-0.052$, $0.119$] & $0.020$ [$-0.038$, $0.077$] \\
TFIM & 6 & 10 & $+0.077$ & $0.052$ [$-0.030$, $0.139$] & $0.024$ [$-0.030$, $0.076$] \\
TFIM & 8 & 4  & $-0.007$ & $0.025$ [$-0.057$, $0.102$] & $\mathbf{-0.032}$ [$-0.060$, $-0.002$] \\
TFIM & 8 & 6  & $+0.040$ & $0.009$ [$-0.090$, $0.094$] & $0.030$ [$-0.020$, $0.081$] \\
TFIM & 8 & 8  & $+0.105$ & $0.040$ [$-0.039$, $0.113$] & $\mathbf{0.065}$ [$0.009$, $0.130$] \\
TFIM & 8 & 10 & $+0.119$ & $0.054$ [$-0.017$, $0.124$] & $\mathbf{0.065}$ [$0.009$, $0.119$] \\
\midrule
LFIM & 4 & 4  & $+0.069$ & $0.048$ [$-0.056$, $0.146$] & $0.021$ [$-0.021$, $0.062$] \\
LFIM & 4 & 6  & $+0.109$ & $0.046$ [$-0.067$, $0.170$] & $\mathbf{0.062}$ [$0.010$, $0.116$] \\
LFIM & 4 & 8  & $+0.105$ & $0.032$ [$-0.073$, $0.165$] & $\mathbf{0.073}$ [$0.017$, $0.129$] \\
LFIM & 4 & 10 & $+0.098$ & $0.022$ [$-0.096$, $0.142$] & $\mathbf{0.076}$ [$0.020$, $0.138$] \\
LFIM & 6 & 4  & $+0.015$ & $0.022$ [$-0.068$, $0.111$] & $-0.007$ [$-0.035$, $0.021$] \\
LFIM & 6 & 6  & $+0.085$ & $0.074$ [$-0.009$, $0.152$] & $0.011$ [$-0.025$, $0.057$] \\
LFIM & 6 & 8  & $+0.133$ & $\mathbf{0.105}$ [$0.033$, $0.183$] & $0.027$ [$-0.019$, $0.080$] \\
LFIM & 6 & 10 & $+0.144$ & $\mathbf{0.114}$ [$0.029$, $0.197$] & $0.030$ [$-0.018$, $0.079$] \\
LFIM & 8 & 4  & $-0.035$ & $-0.023$ [$-0.104$, $0.056$] & $-0.012$ [$-0.046$, $0.024$] \\
LFIM & 8 & 6  & $+0.096$ & $\mathbf{0.085}$ [$0.013$, $0.158$] & $0.011$ [$-0.019$, $0.041$] \\
LFIM & 8 & 8  & $+0.143$ & $\mathbf{0.104}$ [$0.047$, $0.163$] & $\mathbf{0.039}$ [$0.000$, $0.079$] \\
LFIM & 8 & 10 & $+0.168$ & $\mathbf{0.115}$ [$0.061$, $0.174$] & $\mathbf{0.053}$ [$0.008$, $0.104$] \\
\bottomrule
\end{tabular}
\end{table}

For HEA, no systematic coupling is detected in either channel. The
deviations $F - 1$ stay within $\pm 0.035$ in the TFIM and
$\pm 0.019$ in the LFIM, and all within-seed Pearson correlation
intervals contain zero. Isolated between-seed flags appear in one
condition per Hamiltonian, with inconsistent signs and, in the LFIM
case, a normalized covariance contribution of only $0.0015$, so we
treat them as marginal and rest the assessment on the additive
covariance decomposition.
Together with the flat organization factor of the previous
subsections, this places HEA close to the factorized random-sign
predictions at the aggregate second-moment and coupling level. The
scope of these predictions matters. The conditional random-sign
model predicts the two quantities
$\mathbb{E}[B_{\mathrm{eff},k}^2] = 1$ and $F = 1$, and both are
consistent with the HEA data, whereas the activity scale $Q_k$ is
set by the magnitude profile that the model conditions on and is
not predicted by it. The strong decay of $\mathbb{E}[Q_k]$
alongside two model-consistent factors is therefore precisely the
configuration in which the second-moment scaling is carried by
activity rather than by any change in sign organization.

For HVA, a weak coupling emerges toward larger systems. In the
TFIM, $F - 1$ reaches about $0.12$ at $n = 10$ and depth $8$,
carried by both covariance channels with neither consistently
dominating. At the smallest systems some conditions show weakly
negative coupling with intervals excluding zero, so the tested
range contains an observed sign change from negative to positive,
which we report as such rather than as a formally established
transition.

The coupling is where the two Hamiltonians differ most clearly. In
the LFIM, $F$ reaches $1.168$, the confidence interval excludes $1$
in five of the twelve conditions, and at depth $8$ the fitted
$F$ slope interval excludes zero. The channel composition also
shows a depth-associated pattern: the coupling is
between-seed-dominated at depth $4$, within-seed-dominated at depth
$6$, and carried by both channels at depth $8$. We report this as
observed structure of the coupling, without attributing it to the
added longitudinal sector, since the HVA parameter structure also
differs between the two models. 
In the share accounting of the previous subsection the coupling
remains a minor component, but it is the clearest quantitative
difference between the two Hamiltonians, and it identifies the LFIM
HVA as the setting where organization and activity are most
strongly intertwined.

\paragraph{From the second moment to the variance.}
All decomposition statements above are made at the level of $M_2$.
Their transfer to the variance object of Eq.~\eqref{eq:var_def} is
controlled by the mean-gradient correction
$\mathbb{E}_k[\mu_k^2] / M_2$, which we estimated with the
bias-corrected procedure of Sec.~\ref{ss:statistics}, 
since the naive plug-in values of $1.5$--$4.4\%$ are largely
attributable to the $O(1/S)$ finite-sample bias. After correction,
the estimates are small and all $24$ confidence intervals per
Hamiltonian, one for each of the twelve $(n,d)$ conditions and each
ansatz family, contain zero. The mean-gradient correction
$\mathbb{E}_k[\mu_k^2] / M_2$ is therefore consistent with zero
over the tested range, and the second-moment decomposition carries
over approximately to the variance. In summary, the decomposition
answers the question posed at the start of this section: in HEA the
$n$-dependence of the gradient second moment is carried almost
entirely by the decay of termwise activity, with organization and
coupling consistent with their random-sign reference values at the
aggregate level, whereas in HVA activity and organization
contribute comparably, with a weak coupling that is quantitatively
stronger in the LFIM.

\section{Microscopic Sign-Alignment Structure}
\label{s:interpret}
\label{s:microscopic}

The decomposition of the previous section resolves the HVA
second-moment growth into comparable activity and organization
contributions together with a smaller coupling component. This
section examines what that organization looks like at the
microscopic level of sign alignment between Hamiltonian-term
contributions.
The object of study is the within-circuit, parameter-averaged
sign-alignment matrix among termwise gradient contributions, whose
entries measure the tendency of two Hamiltonian terms to contribute
with equal or opposite signs, estimated over $30$ seeds per
condition, with the estimand, the zero-handling rule, and the
random-sign null construction specified in
Sec.~\ref{ss:sign_alignment}.
One property of the reference
distribution matters for everything that follows: the conditional
random-sign null preserves the zero pattern and removes both
pairwise sign alignment and marginal sign imbalance, so all
comparisons against it are consistency tests against the strict 
random-sign description, with literally independent and symmetric signs, 
rather than tests of pairwise correlation alone, 
and this limitation applies symmetrically to every finding
below, for both ansatz families.

\subsection{Sector-Level Alignment in HVA}

\begin{figure}[!htbp]
    \centering
    \includegraphics[width=1.0\textwidth]{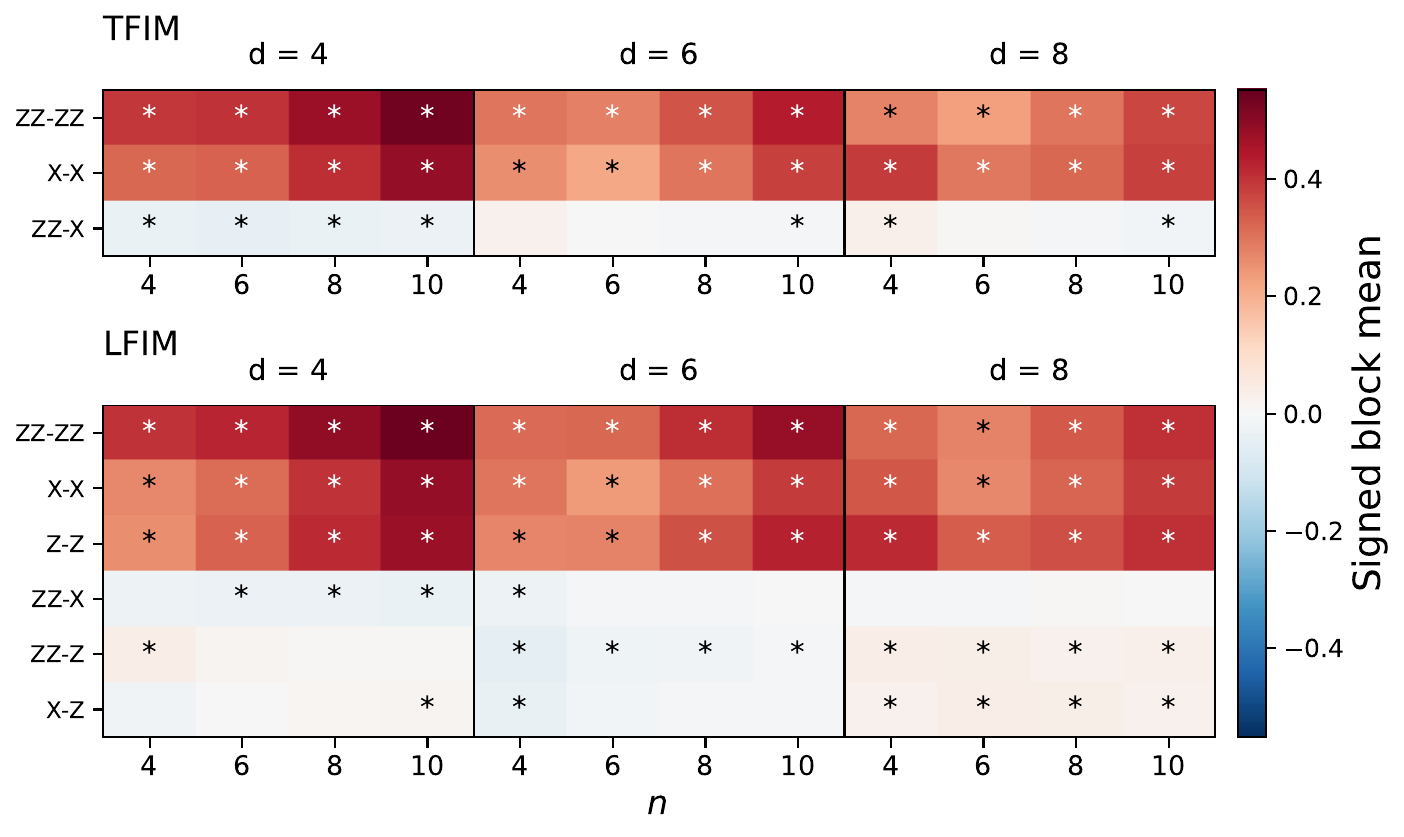}
    \caption{Blockwise signed alignment structure of the HVA
    for the TFIM (top) and the LFIM (bottom). Each cell shows the
    signed sector-block mean of the parameter-averaged
    sign-alignment matrix for one $(n, d)$ condition, with rows
    labeled by sector pairs and columns grouped by circuit depth.
    Asterisks mark cells whose observed value lies outside the
    central 95\% band of the conditional random-sign null, on either
    side, with the sign of the deviation carried by the color. The
    within-sector rows show systematic positive alignment across the
    tested conditions, strengthening toward larger $n$, while the
    cross-sector rows remain substantially weaker and
    include the depth-associated direction change of the $ZZ$--$Z$
    block described in the text. The estimand, the zero-handling
    rule, and the null construction are specified in
    Sec.~\ref{ss:sign_alignment}.}
    \label{fig:sign_alignment}
\end{figure}

For the HVA, the alignment structure is strongly organized along the
Hamiltonian sectors. As shown in
Figure~\ref{fig:sign_alignment}, the within-sector signed alignment
lies above the null in every tested condition: in the TFIM, both the
$ZZ$--$ZZ$ and the $X$--$X$ block means exceed the null band in all
twelve $(n,d)$ conditions, reaching $0.53$ and $0.48$ at $n = 10$
and depth $4$, and in the LFIM the same holds for all three within-sector blocks,
including the $Z$--$Z$ block of the added longitudinal sector. The
pattern is systematic positive alignment across the tested
conditions, with an overall strengthening toward larger $n$. 
The same tendency appears in the global spectral statistic: the
leading eigenvalue of the alignment matrix, reported relative to
its matched-null median, lies between $1.45$ and $2.35$ in the TFIM
and between $1.77$ and $3.06$ in the LFIM across the tested
conditions, with the larger values occurring at larger $n$, in
parallel with the growth of the organization factor in
Section~\ref{s:decomposition}.

Cross-sector alignment behaves differently. Its absolute magnitude
exceeds the block-specific null in every condition for both
Hamiltonians, so cross-sector contributions are not simply
unstructured, but the signed means are an order of magnitude below
the within-sector values. In the TFIM they show a weak negative
signed alignment relative to the conditional random-sign null,
resolved at depth $4$ and at the largest system sizes. In the LFIM
the cross-sector signed deviations show a depth-associated pattern:
the $ZZ$--$X$ direction reproduces the weak negative tendency of
the TFIM at depth $4$, while the $ZZ$--$Z$ block changes direction
with depth, lying below the null at depth $6$ and above it at depth
$8$.  We report these as depth-associated changes in the direction of
cross-sector signed alignment relative to the null. As a physical
reading, the weak negative tendency observed for the ZZ--X pair,
whose two sectors do not commute, is consistent with competition
between noncommuting Hamiltonian sectors. The ZZ--Z pair consists
of mutually commuting sectors, so its depth-associated direction
change is not covered by this reading and is recorded simply as an
observed pattern. Because the null removes both pairwise sign
alignment and marginal sign imbalance, these signed deviations
cannot be interpreted specifically as measured pairwise
anti-correlation. Separating pairwise association from marginal
sign bias requires a refined null construction and is left for
future work.

Taken together, the HVA pattern is sector-structured: strong
same-sign alignment within each Hamiltonian sector, weak and partly
sign-changing structure across sectors. This provides microscopic
support for reading the organization growth of
Section~\ref{s:decomposition} as sector-aligned sign organization,
rather than direct verification, since the null does not decompose
the within-sector signal into pairwise-correlation and
marginal-bias components.

\subsection{Approximate Random-Sign Structure in HEA}

For the HEA, the aggregate results of
Section~\ref{s:decomposition} left open how literally the
random-sign description holds at the microscopic level. The
alignment analysis gives a precise answer: a small but reproducible
excess over the strict null exists. The mean absolute alignment and
the leading eigenvalue of the alignment matrix lie above the
conditional random-sign null in all twelve $(n,d)$ conditions in both
Hamiltonians, with eigenvalue ratios to the null median between
$1.04$ and $1.16$ in the TFIM and between $1.06$ and $1.30$ in the
LFIM. Even the tightest case, a mean absolute alignment of $0.0597$
against a null 95th percentile of $0.0569$ at $n = 10$ and depth
$8$ in the TFIM, persists at $1000$ null replicates. 
The HEA statistics are essentially insensitive to the
zero-threshold choice across the swept range, with changes at most
of order $10^{-5}$ in the mean absolute alignment, so the excess is
not an artifact of the numerical-zero handling.
The two statistics are
strongly dependent summaries of the same alignment matrix, and the
twelve conditions share the ansatz family, so we do not count these
as independent exceedances. The appropriate conclusion is that a
reproducible, same-direction pattern appears across all tested
conditions, which is sufficient to conclude that HEA is not fully
consistent with the strict conditional random-sign null.

This residual structure does not accumulate into net signed
organization. The signed sector-block means remain near zero,
between $-0.017$ and $+0.028$ in the TFIM, and the two-sided
signed-block flags are scattered, $6$ of $36$ in the TFIM and $15$
of $72$ in the LFIM, with inconsistent directions, so no individual
flag should be overread. The excess is also sector-nonspecific: the
absolute alignment exceeds its null in every sector block, without
any sector-selective signed pattern, which is the sharpest contrast
with the sector-aligned HVA structure of the previous subsection.
The size of the excess shows an overall tendency to weaken toward
larger systems at fixed depth, while its dependence on depth is not
monotonic.

These observations combine into a three-layer description of the
HEA sign structure. At the aggregate level the organization factor
is consistent with the exact random-sign value, at the microscopic
alignment level a small reproducible excess over the strict null exists, 
and at the signed sector level the means remain near zero.
The strongest interpretation the data support is that a weak
residual sign structure is present, compatible with weak mixed-sign
alignment, with marginal sign imbalance, or with both, possibilities
the null construction cannot distinguish, and that this structure
does not build up into the kind of systematic sign organization that
would shift the organization factor.
This interpretation is consistent with the role of the random-sign
model introduced in Section~\ref{s:framework} as an aggregate
reference for unstructured cancellation. Agreement with the
random-sign baseline at the level of the organization factor does
not require the microscopic signs themselves to be independent and
symmetric. The HEA therefore illustrates how weak residual
microscopic structure can coexist with aggregate organization that
remains close to the random-sign reference.

\paragraph{Relation to generator--Hamiltonian alignment.}
The contrast between the two microscopic patterns admits a natural
structural reading. The HVA layer generators are drawn from the
operator sectors of the target Hamiltonian, and the within-sector
alignment documented above is organized along the sectors
from which those generators are drawn, whereas the HEA, whose
generators are not tied to the Hamiltonian decomposition, shows only
weak and sector-nonspecific residual structure. This correspondence
is consistent with the view, discussed from the Lie-algebraic side
in Section~\ref{ss:lie}, that generator--Hamiltonian alignment
supports the kind of sign organization that the organization factor
measures. Two limitations keep this at the level of an
interpretation. The comparison design does not isolate generator
alignment as a single variable, since the two families also differ
in parameter count, parameter sharing, and gate structure. And the
alignment analysis itself is a consistency test whose null does not
separate pairwise correlation from marginal sign imbalance. We
therefore read generator--Hamiltonian alignment as consistent with
the observed sector structure and as a working hypothesis for
ansatz design, not as an established mechanism.

\section{Discussion}
\label{s:discuss}

This work set out to resolve the internal composition of
BP-relevant gradient suppression at the level of
Hamiltonian-term contributions. The preceding sections developed
the termwise decomposition, measured its three factors across two
ansatz families and two related Ising Hamiltonians, and
characterized the resulting sign organization at the microscopic
level. We now discuss what these measurements imply for the
decomposition, the random-sign reference model, and the scope and
limitations of the diagnostics.

\paragraph{What the decomposition adds.}
The central analytical device of this work is a lossless
decomposition of the gradient second moment into an activity scale,
an organization factor, and their coupling, linked to the standard
variance-based language of BP theory through the variance bridge
and the mean-gradient check, which shows that the second-moment
conclusions carry over approximately to the variance over the
tested range. What this adds to the standard formulation is
separability: the three factors can be measured separately on the
same data, and the conditional random-sign model makes falsifiable
predictions for two of them.
The value of the framework therefore lies not in the
identity itself but in what the separate measurements reveal about
where the $n$-dependence of the gradient signal actually resides.

\paragraph{Where the scaling resides.}
The measurements answer this question differently for the two
ansatz families, at two levels of resolution. For the HEA, the
second-moment suppression over the tested range is carried almost
entirely by the decay of termwise activity, while the organization
factor remains near the random-sign reference without systematic
scaling and the coupling factor remains statistically consistent
with its random-sign reference. For the HVA, activity and
organization grow comparably, and the coexistence of larger
$N_{\mathrm{eff}}$ with larger $B_{\mathrm{eff}}$ rules out simple
sparsification as the origin of the stronger organization observed
in the HVA. The coupling factor is a minor component of the slope
accounting for both families, yet it provides the clearest
quantitative difference between the two Hamiltonians, with
departures from $F=1$ that are stronger in the LFIM.
At the microscopic level, the sign-alignment analysis sharpens both
pictures: the HVA organization is structured along the Hamiltonian
sectors, while the HEA combines aggregate organization close to the
random-sign reference with a weak residual sign structure that does
not accumulate into net signed organization, illustrating that
aggregate agreement with the reference does not require the
microscopic signs themselves to be independent and symmetric.

\paragraph{Scope of the decomposition and the random-sign model.}
The random-sign model also bounds the scaling contribution that
unstructured signed cancellation can make.
Under the conditional random-sign model, the expected gradient second moment
is reduced, in conditional expectation at fixed magnitudes, by
exactly the factor $N_{\mathrm{eff},k}$ relative to a fully
sign-aligned reference with the same magnitude profile, since the
aligned reference attains $B_{\mathrm{eff},k}^2 = N_{\mathrm{eff},k}$
while the random-sign value is $1$ by Eq.~\eqref{eq:B2_baseline}.
The effective term count is bounded by the number of Hamiltonian
terms, which is linear in $n$ for the Ising chains studied here and
polynomial for Hamiltonians with polynomially many terms more
generally. Random-sign cancellation can therefore contribute at
most a polynomially bounded relative suppression, and within this
setting an exponential decay of the gradient second moment must be
carried by the activity scale. 
The HEA measurements are consistent with this picture: activity decays
with system size, while the organization factor remains near the
random-sign reference without systematic scaling and the coupling
factor remains statistically consistent with its reference. Thus, over
the tested range, the BP-relevant suppression in the HEA is associated
with decaying activity rather than progressively deepening cancellation.
A qualitatively stronger cancellation route
remains possible in principle: organized anti-correlation among termwise
contributions could drive $\mathbb{E}[B_{\mathrm{eff},k}^2]$ toward
zero and produce suppression beyond the random-sign bound. We do
not observe this route in either family. The most direct evidence
is aggregate: the organization factor remains consistent with the
random-sign value in the HEA and grows in the HVA, so neither
family trends toward vanishing organization. The microscopic
analysis reinforces this, since the negative cross-sector
deviations in the HVA remain an order of magnitude below the
within-sector alignments and the HEA residual structure does not
accumulate into net signed organization. The anti-correlation route
nevertheless identifies the regime in which cancellation itself,
rather than activity, could in principle carry BP scaling.

\paragraph{Role and limits of the diagnostics.}
The quantities studied here are descriptive statistics of gradient
structure at random initialization. The group-level factors
$\mathbb{E}[Q_k]$, $\mathbb{E}[B_{\mathrm{eff},k}^2]$, and $F$,
together with the underlying parameter-level diagnostics $R_k$,
$N_{\mathrm{eff},k}$, $B_{\mathrm{eff},k}$, and $Q_k$, 
characterize how the gradient
second moment is composed over the tested range, and their
interpretation is inherently joint: the second moment is the
product of the three group-level factors, so no single factor
determines the gradient scale, and an elevated organization factor
by itself is not a trainability mechanism.
For the same reason, the diagnostics
are not surrogates for optimization performance. They do not by
themselves predict convergence speed or final objective values,
which also depend on how update directions align with the
objective, on curvature, and on the geometry of the optimization
path. The framework should therefore be read as a diagnostic
account of where gradient signal resides and how it is organized at
initialization, not as a predictor of optimization success.

\paragraph{Outlook.}
Several directions follow from this work. 
First, the decomposition should be applied beyond HEA and HVA to
broader classes of structured ansatzes, to test how generally the
two profiles observed here, activity-carried suppression on one
side and joint activity-organization growth on the other, describe
the gradient scaling that is relevant to trainability.
Second, the generator--Hamiltonian
alignment reading, which the present data support as a working
hypothesis rather than an established mechanism, should be
developed into an explicit design question: which forms of
alignment support sector-structured sign organization, and under
which circuit modifications that organization degrades. Third, the
present study covers two related Ising Hamiltonians at statevector
scale, so it remains to examine whether the same decomposition
profiles persist for other Hamiltonian families, larger systems,
and sampled rather than exact expectation values. Fourth, the
sign-alignment analysis motivates a refined null construction: a
marginal-preserving permutation null would separate pairwise sign
correlation from marginal sign imbalance, which the conditional
random-sign null used here removes jointly, and would allow the
weak cross-sector structure observed in the HVA to be attributed
to one component or the other. 
Together with the anti-correlation regime outlined above, these
directions turn the decomposition from a diagnostic of two ansatz
families into a broader account of how activity and organization
shape trainability-relevant gradient structure.

\section{Conclusion}
\label{s:conc}

In this paper, we developed a diagnostic framework that decomposes
the gradient statistics of variational quantum circuits into the
signed contributions of individual Hamiltonian terms. The central
object is an exact decomposition of the gradient second moment into
an activity scale, an organization factor, and their coupling,
linked to the standard variance-based description of BPs through
the variance bridge and an explicit mean-gradient check. The
conditional random-sign model supplies falsifiable reference
predictions for the organization and coupling factors. We measured
all three factors for hardware-efficient and Hamiltonian-variational
ansatz families on two related Ising Hamiltonians, and further
examined the microscopic sign-alignment structure underlying the
organization factor.

The measurements give the two ansatz families qualitatively
different profiles. In the hardware-efficient family, the
finite-size suppression of the gradient second moment is carried
almost entirely by the decay of termwise activity, while the
organization and coupling factors remain statistically consistent
with their random-sign reference values at the aggregate level, and
the microscopic analysis adds only a weak residual sign structure
that does not accumulate into net signed organization. In the
Hamiltonian-variational family, activity and organization grow
comparably with system size, the growth in organization reflects
broad term participation with sector-structured sign alignment
rather than a reduction in the number of active terms, and the
gradient second moment increases over the tested range. Both
profiles are reproduced across the two Ising Hamiltonians, with the
clearest quantitative difference between them appearing in the
coupling factor, whose departures from unity are stronger in the
longitudinal-field model.

Under the conditional random-sign model, and for Hamiltonians whose
term decomposition contains polynomially many terms, random-sign
cancellation produces at most a polynomially bounded suppression
relative to a fully sign-aligned reference with the same magnitude
profile. An exponential decay of the gradient second moment within
this setting must therefore be carried by the activity scale, and
the hardware-efficient measurements are consistent with this
division of roles. Stronger suppression through organized
anti-correlation remains possible outside this setting and is not
observed here. The value of the decomposition is that it makes the
composition of gradient suppression measurable: it separates what
the magnitudes do from how the signs are organized, makes the
random-sign description falsifiable at both the aggregate and
microscopic levels, and locates each ansatz family within that
decomposition. We expect this view to provide a useful starting
point for analyzing gradient structure in other ansatz families and
Hamiltonians, and for treating activity and organization as
distinct considerations in future ansatz design.

\section{Methods}
\label{s:methods}

This section describes the Hamiltonians, circuit constructions,
simulation procedures, and statistical aggregation protocols used
in the study. It then specifies the termwise diagnostics and the
second-moment decomposition as formal estimands, the estimation and
uncertainty-quantification procedures for all reported quantities,
and the sign-alignment analysis, with the TFIM and the
LFIM treated throughout by a common analysis pipeline.

\subsection{Hamiltonian and Circuit Definitions}
\label{ss:circuit_definitions}

We consider the TFIM with open boundary conditions as the primary Hamiltonian throughout the study. For an $n$-qubit system, the Hamiltonian is defined as
\begin{equation}
H
=
-\sum_{i=1}^{n-1} Z_i Z_{i+1}
+
h\sum_{i=1}^{n} X_i,
\end{equation}
where $Z_i$ and $X_i$ denote the Pauli-$Z$ and Pauli-$X$ operators acting on qubit $i$, and $h$ is the transverse-field strength. We set $h = 1$ throughout this study. The first term represents nearest-neighbor Ising interactions along the chain, while the second term represents the local transverse field. We use open boundary conditions throughout, so the interaction term includes only pairs $(i,i+1)$ for $i=1,\dots,n-1$ and does not include a wraparound coupling between the first and last qubits.

To probe the robustness of our findings beyond a single Hamiltonian,
we additionally study the LFIM, a
minimal extension of the TFIM obtained by adding a uniform
longitudinal field. For an $n$-qubit chain with open boundary
conditions, the Hamiltonian is
\begin{equation}
\label{eq:lfim_hamiltonian}
H_{\mathrm{LFIM}}
=
-\sum_{i=1}^{n-1} Z_i Z_{i+1}
+
h_x \sum_{i=1}^{n} X_i
+
h_z \sum_{i=1}^{n} Z_i,
\end{equation}
where $h_x$ and $h_z$ denote the transverse-field and
longitudinal-field strengths. We set $h_x = 1.0$ and $h_z = 0.5$
throughout. The longitudinal field breaks the $\mathbb{Z}_2$
spin-flip symmetry of the TFIM and adds a third operator sector of
diagonal single-qubit terms. The LFIM therefore provides a natural
first test of whether the decomposition pattern observed for the
TFIM persists under a modified sector structure.

We study two ansatz families with contrasting structural properties.
The HEA is composed of repeated layers of local single-qubit
rotations followed by a generic entangling pattern. Concretely, each
HEA layer applies trainable $R_y$ and $R_z$ rotations to every qubit
and then applies a nearest-neighbor ring of CNOT gates. The same HEA
construction is used for both Hamiltonians. By contrast, the HVA is
constructed directly from the operator sectors appearing in the
target Hamiltonian. For the TFIM, each HVA layer consists of a field
block generated by $\sum_i X_i$, implemented as single-qubit
$R_x(2\beta)$ rotations, followed by an interaction block generated
by $\sum_{i=1}^{n-1} Z_i Z_{i+1}$, implemented as nearest-neighbor
$R_{ZZ}(2\gamma)$ gates. The blocks are generated by the sector operators 
themselves, and any sign of the corresponding Hamiltonian coefficient 
is absorbed into the trainable angle, whose initialization is 
symmetric about zero. 
For the LFIM, each HVA layer contains one
additional block that mirrors the added Hamiltonian sector: the
transverse-field block $R_x(2\beta_x)$ and the interaction block
$R_{ZZ}(2\gamma_{zz})$ are followed by a longitudinal-field block
generated by $\sum_i Z_i$, implemented as single-qubit
$R_z(2\beta_z)$ rotations, so that a single layer carries the three
parameters $(\beta_x, \gamma_{zz}, \beta_z)$. Thus, HEA represents a
generic expressive circuit with weak problem alignment, whereas HVA
represents a problem-informed circuit whose generators remain
directly matched to the Hamiltonian structure.

For reference in the analyses that follow, we record the term and
parameter counts of the two models. The TFIM contains $M = 2n - 1$
Hamiltonian terms, namely $n-1$ nearest-neighbor $ZZ$ terms and $n$
transverse-field $X$ terms. The LFIM contains $M = 3n - 1$ terms,
adding $n$ longitudinal-field $Z$ terms to the same $ZZ$ and $X$
sectors. Throughout the termwise analyses we keep a fixed term
ordering that groups the terms by sector, which the sector-resolved
sign-alignment analysis of Sec.~\ref{ss:sign_alignment} relies on.
On the circuit side, the HEA carries $2nd$ trainable parameters at
depth $d$, corresponding to the $R_y$ and $R_z$ rotations on each of
the $n$ qubits in each of the $d$ layers. The HVA carries $2d$
parameters for the TFIM and $3d$ parameters for the LFIM, since
every layer contributes one parameter per Hamiltonian sector.

\subsection{Statevector Simulation and Gradient Evaluation}

All experiments were performed by exact statevector simulation. For each choice of system size, circuit depth, ansatz family, and random initialization seed, we constructed the full $2^n$-dimensional quantum state and evaluated the cost function defined in Eq.~\eqref{eq:cost} directly from the resulting statevector. This exact simulation setting introduced no sampling noise or measurement shot error, and was adopted in order to isolate the termwise gradient structure without additional stochastic variation from finite-shot estimation.

Gradient evaluation was performed differently for the two ansatz families. For the HEA, gradients were computed using the exact parameter-shift rule~\cite{schuld:2019:pra}, since each trainable parameter enters through a standard single-qubit rotation generated by a simple Pauli operator. Concretely, for a parameter $\theta_k$, we evaluated
\begin{equation}
\frac{\partial \langle H \rangle}{\partial \theta_k}
=
\frac{1}{2}
\left[
\langle H \rangle_{\theta_k+\pi/2}
-
\langle H \rangle_{\theta_k-\pi/2}
\right].
\end{equation}
This makes analytic shift-based differentiation straightforward in the statevector setting.

For the HVA, by contrast, gradients were evaluated throughout by central finite differences (FD). Specifically, with $\varepsilon=10^{-5}$, we used
\begin{equation}
\frac{\partial \langle H \rangle}{\partial \theta_k}
\approx
\frac{
\langle H \rangle_{\theta_k+\varepsilon}
-
\langle H \rangle_{\theta_k-\varepsilon}
}{
2\varepsilon
}.
\end{equation}
This choice was made for practical and methodological consistency: HVA parameters are shared across problem-informed operator blocks, and the central FD procedure provided a uniform gradient-evaluation pipeline for both full-Hamiltonian and termwise contributions.

As a sensitivity check, we repeated the diagnostic computation in representative small and large system settings using nearby step sizes, $\varepsilon=10^{-4}$ and $10^{-6}$. The qualitative HEA--HVA ordering of $N_{\mathrm{eff},k}$ and $B_{\mathrm{eff},k}$ remained unchanged, indicating that the observed diagnostic ordering is not an artifact of the FD step size.

\subsection{Termwise Decomposition and Diagnostic Computation}
\label{ss:termwise_methods}

For each parameter $\theta_k$, we used the termwise decomposition in Eq.~\eqref{eq:termwise_grad} to compute Hamiltonian-term contributions separately. Specifically, each $a_{\alpha,k}$ was obtained by applying the same differentiation rule used for the full gradient, but with the full Hamiltonian $H$ replaced by the corresponding single Hamiltonian term $H_\alpha$. In this way, the raw termwise contributions $\{a_{\alpha,k}\}$ for each parameter $k$ provide a complete term-resolved representation of the gradient.

From these termwise contributions, we computed the diagnostics defined in Sections~\ref{s:framework} and~\ref{s:var_bridge}, namely $R_k$, $N_{\mathrm{eff},k}$, $B_{\mathrm{eff},k}$, and $Q_k$. 
These quantities characterize the activity and sign-organization
structure of the termwise gradient contributions and provide the
inputs to the second-moment decomposition used in the main text.

At the level of individual parameters, these diagnostics are tied
together by the exact pointwise identity
$(\partial_k \langle H\rangle)^2 = B_{\mathrm{eff},k}^2\, Q_k$ of
Eq.~\eqref{eq:variance_bridge_identity}, which holds algebraically
for every parameter and every run. We verified this identity directly
in the recorded data. Across all $(n,d)$ settings, both ansatz
families, and both Hamiltonians, the reconstructed product
$B_{\mathrm{eff},k}^2 Q_k$ matched the directly computed squared
gradient with maximum absolute residuals of order $10^{-13}$ and
median residuals at the $10^{-17}$ level. This confirms the internal
consistency of the termwise decomposition with the full-gradient
numerical evaluation, since both are computed with the same
differentiation rule.

For ensemble-level analysis we group runs by system size, circuit
depth, and ansatz family. Within a group, we write
$\mathbb{E}_s[\cdot]$ for the average over initialization seeds at a
fixed parameter index and $\mathbb{E}_k[\cdot]$ for the subsequent
average over trainable parameters, and we abbreviate the combined
average as $\mathbb{E}[\cdot] = \mathbb{E}_k[\mathbb{E}_s[\cdot]]$.
The estimator and its uncertainty quantification are specified in
Sec.~\ref{ss:statistics}. Two aggregate objects must be
distinguished. The gradient second moment
\begin{equation}
M_2 = \mathbb{E}\big[(\partial_k \langle H\rangle)^2\big]
\label{eq:M2_def}
\end{equation}
is the quantity that our decomposition addresses directly. The
variance object relevant to barren-plateau analysis is the
per-parameter variance over initializations, averaged over
parameters. Writing $\mu_k = \mathbb{E}_s[\partial_k \langle
H\rangle]$ for the seed-mean gradient of parameter $k$, this object
is
\begin{equation}
\mathbb{E}_k\big[\mathrm{Var}_s(\partial_k \langle H\rangle)\big]
=
M_2 - \mathbb{E}_k\big[\mu_k^2\big],
\label{eq:var_def}
\end{equation}
which coincides with $M_2$ exactly when $\mu_k = 0$ for every
parameter. In this work all decomposition statements are made at the
level of $M_2$, and the size of the mean-gradient correction
$\mathbb{E}_k[\mu_k^2] / M_2$ separating the two objects is estimated
with a dedicated bias-corrected procedure described in
Sec.~\ref{ss:statistics}.

Taking expectations of the pointwise identity yields
$M_2 = \mathbb{E}[B_{\mathrm{eff},k}^2 Q_k]$, which we analyze
through the three-factor decomposition
\begin{equation}
M_2
=
\mathbb{E}[B_{\mathrm{eff},k}^2]\,
\mathbb{E}[Q_k]\,
F,
\qquad
F
=
\frac{\mathbb{E}[B_{\mathrm{eff},k}^2 Q_k]}
     {\mathbb{E}[B_{\mathrm{eff},k}^2]\,\mathbb{E}[Q_k]}.
\label{eq:three_factor}
\end{equation}
The factor $F$ is the coupling factor introduced in
the variance-bridge analysis of Sec.~\ref{s:var_bridge}, now treated
as a named estimand on the same footing as
$\mathbb{E}[B_{\mathrm{eff},k}^2]$ and $\mathbb{E}[Q_k]$. It
quantifies the coupling between sign organization and activity, and
$F = 1$ corresponds to exact factorization. Because the decomposition
is multiplicative, changes across system sizes add on a logarithmic
scale,
\begin{equation}
\Delta \log M_2
=
\Delta \log \mathbb{E}[B_{\mathrm{eff},k}^2]
+
\Delta \log \mathbb{E}[Q_k]
+
\Delta \log F,
\label{eq:log_decomposition}
\end{equation}
which is the form used for the slope-share analysis of
Sec.~\ref{s:decomposition}.

\subsection{Experimental Settings and Statistical Aggregation}
\label{ss:aggregation_methods}

The experiments were carried out over system sizes $n \in \{4,6,8,10\}$ and
circuit depths $d \in \{4,6,8\}$ for both HEA and HVA, and for both
Hamiltonians. For each $(n,d)$ setting and each ansatz family, 30 random
initialization seeds were used, and all diagnostics were computed separately for
every run before aggregation. This design allowed us to compare the two ansatz
families over a common grid of problem sizes and circuit depths while also
assessing the stability of the observed diagnostic patterns across random
initial conditions.

For statistical aggregation, runs were grouped by $(n,d,\text{variant})$ for
each Hamiltonian. Within each group, diagnostics were first averaged over
parameters within each run, and we reported the mean of
these run-level averages across the 30 seeds for $R_k$, $N_{\mathrm{eff},k}$,
and $B_{\mathrm{eff},k}$. Because the design is balanced, the resulting mean
values coincide with the two-level average $\mathbb{E}_k[\mathbb{E}_s[\cdot]]$
introduced in Sec.~\ref{ss:termwise_methods}, while the reported standard
deviations quantify seed-to-seed variability rather than pooled parameter-level
dispersion. 

For decomposition and coupling analysis, we computed both the exact product
moment and its factorized counterpart at the group level.  Specifically, for
each $(n,d,\mathrm{variant})$ group, we evaluated
$\mathbb{E}[B_{\mathrm{eff},k}^2Q_k]$ directly from the pooled parameter-level
data and compared it with $\mathbb{E}[B_{\mathrm{eff},k}^2]\mathbb{E}[Q_k]$.  To
quantify the quality of factorization, we reported the coupling factor $F$
defined in Eq.~\eqref{eq:three_factor}, together with the empirical correlation
$\mathrm{corr}(B_{\mathrm{eff},k}^2,Q_k)$. A value of $F$ close to 1 indicates
that the factorization is accurate for the group, and deviations of $F$ from 1
directly quantify coupling between $B_{\mathrm{eff},k}^2$ and $Q_k$. Because the
random-sign model of Section~\ref{s:framework} predicts $F = 1$, sustained
deviations can additionally be read as structure beyond that reference model.
The empirical correlation was computed and used as complementary interpretive 
support rather than as a primary coupling measure.

\subsection{Statistical Estimation and Uncertainty Quantification}
\label{ss:statistics}

All statistical statements in this work treat the initialization seed
as the independent unit of replication. Within each
$(n,d,\text{variant})$ group and for each Hamiltonian, every
group-level moment entering the decomposition, namely $M_2$,
$\mathbb{E}[B_{\mathrm{eff},k}^2]$, $\mathbb{E}[Q_k]$, and
$\mathbb{E}[B_{\mathrm{eff},k}^2 Q_k]$, is estimated following the
two-level convention of Sec.~\ref{ss:termwise_methods}: for each
parameter the quantity is averaged over the 30 seeds, and the
resulting per-parameter values are then averaged over the trainable
parameters. Because these moments are linear in the underlying
per-run values, the two averages commute, and the same estimates are
obtained from the run-level parameter averages described in
Sec.~\ref{ss:aggregation_methods}. The order becomes essential only
for nonlinear functionals such as the mean-gradient ratio below,
where the seed average must be taken first. Derived quantities such
as the coupling factor $F$ of Eq.~\eqref{eq:three_factor} are
computed as plug-in functions of these moment estimates.

Uncertainty was quantified by a nonparametric bootstrap over seeds.
In each bootstrap iteration, the 30 seeds of a group were resampled
with replacement, and all moments and derived ratios of that group
were recomputed from the same resample, so that quantities compared
within a group inherit a consistent uncertainty assessment. Distinct
groups were resampled independently of one another, matching the
independence of the underlying runs, and quantities that combine
several groups, such as the scaling fits below, were recomputed in
each iteration from the independently resampled groups. We used 5000
iterations for the scaling and decomposition analyses and 1000
iterations for the covariance decomposition of the coupling factor,
whose within-seed and between-seed components are computationally
heavier to recompute. We report 95\% percentile confidence
intervals, taken between the 2.5th and 97.5th percentiles of the
bootstrap distributions.

Scaling with system size was summarized by log-linear fits. For each
depth and ansatz family, the logarithm of a group-level quantity was
regressed on $n$ by least squares, and we report the fitted slope
together with the per-qubit factor $e^{\mathrm{slope}}$. Slope
uncertainty was obtained by refitting on each bootstrap resample. We
describe the observed decay and growth as log-linear over the tested
range $n \in \{4,\dots,10\}$ and make no claim about asymptotic
behavior beyond it. Based on the additive form of
Eq.~\eqref{eq:log_decomposition}, we further report slope shares,
defined as the ratio of each component slope to the slope of
$\log M_2$. These shares describe the composition of the
second-moment log change over the tested range and carry no causal
attribution.

The coupling factor was analyzed through its covariance
representation. With plug-in moments, computed with denominator $N$
rather than $N-1$, the identity
\begin{equation}
F - 1
=
\frac{\mathrm{Cov}\big(B_{\mathrm{eff},k}^2,\, Q_k\big)}
     {\mathbb{E}[B_{\mathrm{eff},k}^2]\,\mathbb{E}[Q_k]}
\label{eq:F_cov}
\end{equation}
holds exactly on the data. Applying the law of total covariance over
seeds splits this normalized covariance into a within-seed component
and a between-seed component, and the two components add exactly to
$F - 1$ under the same plug-in convention. In this representation,
$F = 1$ means factorization of the specific product moment
$\mathbb{E}[B_{\mathrm{eff},k}^2 Q_k]$, equivalently zero normalized
covariance between $B_{\mathrm{eff},k}^2$ and $Q_k$, and does not
mean statistical independence of the two quantities. Pearson
correlations between $B_{\mathrm{eff},k}^2$ and $Q_k$ were computed
alongside this decomposition and are used as complementary
interpretive support, as stated in
Sec.~\ref{ss:aggregation_methods}.

Finally, the mean-gradient correction separating $M_2$ from the
variance object of Eq.~\eqref{eq:var_def} was estimated with an
explicit bias correction. The estimand is the mean-gradient ratio
$\mathbb{E}_k[\mu_k^2] / M_2$. The naive plug-in estimator, which
squares the seed-mean gradient $\bar{g}_k$ of each parameter,
carries an upward bias of order $1/S$ because
$\mathbb{E}[\bar{g}_k^2] = \mu_k^2 + \sigma_k^2 / S$ for seed
variance $\sigma_k^2$ and $S = 30$ seeds. We therefore report the
bias-corrected estimator
\begin{equation}
\widehat{\mathrm{MGR}}
=
\frac{\mathbb{E}_k\!\left[\,\bar{g}_k^2 - s_k^2 / S\,\right]}
     {\widehat{M}_2},
\label{eq:mgr_corrected}
\end{equation}
where $s_k^2$ is the unbiased seed-sample variance of parameter $k$.
Negative values of the corrected numerator were retained without
clipping, since discarding them would reintroduce bias. Within each
bootstrap resample, the resampled seed-sample variance was rescaled
by $S/(S-1)$ so that the subtraction term remains calibrated under
resampling with replacement. Confidence intervals for
$\widehat{\mathrm{MGR}}$ containing zero are read as consistent with
a vanishing mean-gradient correction over the tested range.

\subsection{Sign-Alignment Analysis}
\label{ss:sign_alignment}

To probe the sign structure underlying the organization factor at
the microscopic level, we analyzed the alignment of termwise
gradient signs within each circuit. For every run $s$, the raw
termwise contributions $a_{\alpha,k}$ were converted to sign
patterns $\sigma_{\alpha,k}^{(s)} \in \{-1, 0, +1\}$, where entries
with $|a_{\alpha,k}|$ below a zero threshold were assigned
$\sigma = 0$. From these patterns we formed the sign-alignment
matrix
\begin{equation}
C_{\alpha\beta}^{(s)}
=
\frac{1}{K}\sum_{k=1}^{K}
\sigma_{\alpha,k}^{(s)}\,\sigma_{\beta,k}^{(s)},
\label{eq:sign_alignment_matrix}
\end{equation}
where $K$ is the number of trainable parameters, so that each matrix
entry measures the parameter-averaged tendency of two Hamiltonian
terms to contribute with equal or opposite signs. The 30 seeds
provide independent replicates of $C^{(s)}$. We summarized each
matrix by its mean absolute off-diagonal entry, by its leading
eigenvalue $\lambda_{\max}$, and by sector-resolved block averages,
taken in both signed and absolute form over the term blocks defined
by the sector composition of Sec.~\ref{ss:circuit_definitions}.

The zero threshold requires care because the finite-difference
evaluation leaves analytically vanishing contributions at a roundoff
floor rather than at exact zero. 
Such structural zeros arise for final-layer parameters whose
generators commute with a Hamiltonian sector, since no subsequent
noncommuting evolution intervenes before the measurement. For the
TFIM HVA this affects the final-layer interaction parameter and the
$ZZ$ sector, and for the LFIM HVA the two final-layer diagonal
parameters and the $ZZ$ and $Z$ sectors.
We selected the threshold from the data rather than by convention. 
First, the magnitudes at
the theoretically vanishing positions cluster several orders of
magnitude below all remaining entries, leaving a clear gap. Second,
at a threshold of $10^{-10}$ the resulting fractions of zero-classified
entries match the exact theoretical values, $1/(2d)$ for the TFIM
$ZZ$ sector and $2/(3d)$ for the LFIM $ZZ$ and $Z$ sectors, in every
tested condition. 
Third, all reported statistics and flags are stable across
thresholds from $10^{-10}$ to $10^{-8}$, with the observed alignment
statistics essentially identical and residual variation at or below
the Monte Carlo level of the random-sign null. We therefore fixed
the threshold at $10^{-10}$. At the tighter value $10^{-12}$ the
roundoff-scale entries retain effectively random signs, which
slightly inflates the alignment statistics, so the final threshold
also acts conservatively with respect to the reported alignment.

Observed statistics were referenced against a conditional random-sign null. 
For each null replicate, the signs of all nonzero entries of each run were
replaced by independent symmetric random signs while the zero
pattern was held fixed, and the full statistic pipeline, including
the averaging over seeds, was recomputed. We used 1000 null
replicates per condition. This null removes pairwise sign alignment
and marginal sign imbalance simultaneously, so a deviation from it
indicates a departure from the conditional random-sign model as a
whole rather than pairwise correlation specifically. We therefore
interpret the comparison as a consistency test of the random-sign
description, not as an isolated test of pairwise dependence.

Signed block statistics were flagged when the observed value fell
outside the central 95\% range of the null distribution, a two-sided
criterion, since deviations of either sign are meaningful for signed
quantities. Magnitude-type statistics, namely the mean absolute
off-diagonal entry and the absolute block averages, were assessed
against the upper tail of their null distributions. 
The leading eigenvalue was assessed against the upper tail in the
same way and additionally reported as the ratio of the observed
value to the null median, which normalizes its raw scale relative to
a matched null sharing the same matrix size and zero pattern.
Seed-to-seed uncertainty of the
observed statistics was quantified by the bootstrap of
Sec.~\ref{ss:statistics} with 2000 iterations.

\section*{Code and Data Availability}
The code, analysis scripts, and result data supporting this work are
publicly available at \url{https://github.com/pilsungk/BP-DI}.
This includes the experiment drivers for both Hamiltonians, the full
analysis pipeline, the per-run diagnostic data, and the raw termwise
gradient matrices used in the sign-alignment analysis.

\section*{Acknowledgement}
This work was supported by the National Research Foundation of Korea (NRF) grant funded by the Korea government (MSIT), grant number RS-2026-25477171.



\begin{thebibliography}{10}
\providecommand{\url}[1]{#1}
\csname url@samestyle\endcsname
\providecommand{\newblock}{\relax}
\providecommand{\bibinfo}[2]{#2}
\providecommand{\BIBentrySTDinterwordspacing}{\spaceskip=0pt\relax}
\providecommand{\BIBentryALTinterwordstretchfactor}{4}
\providecommand{\BIBentryALTinterwordspacing}{\spaceskip=\fontdimen2\font plus
\BIBentryALTinterwordstretchfactor\fontdimen3\font minus
  \fontdimen4\font\relax}
\providecommand{\BIBforeignlanguage}[2]{{%
\expandafter\ifx\csname l@#1\endcsname\relax
\typeout{** WARNING: IEEEtran.bst: No hyphenation pattern has been}%
\typeout{** loaded for the language `#1'. Using the pattern for}%
\typeout{** the default language instead.}%
\else
\language=\csname l@#1\endcsname
\fi
#2}}
\providecommand{\BIBdecl}{\relax}
\BIBdecl

\bibitem{peruzzo:2014:vqe}
A.~Peruzzo, J.~McClean, P.~Shadbolt, M.-H. Yung, X.-Q. Zhou, P.~J. Love,
  A.~Aspuru-Guzik, and J.~L. O'Brien, ``{A Variational Eigenvalue Solver on a
  Photonic Quantum Processor},'' \emph{Nature Communications}, vol.~5, no.~1,
  p. 4213, Jul 2014.

\bibitem{farhi:2014:qaoa}
E.~Farhi, J.~Goldstone, and S.~Gutmann, ``{A Quantum Approximate Optimization
  Algorithm},'' 2014.

\bibitem{cerezo:2021:variational}
M.~Cerezo, A.~Arrasmith, R.~Babbush, S.~C. Benjamin, S.~Endo, K.~Fujii, J.~R.
  McClean, K.~Mitarai, X.~Yuan, L.~Cincio, and P.~J. Coles, ``{Variational
  Quantum Algorithms},'' \emph{Nature Reviews Physics}, vol.~3, no.~9, pp.
  625--644, 2021,
  \href{https://doi.org/10.1038/s42254-021-00348-9}{doi:10.1038/s42254-021-00348-9}.

\bibitem{mcclean:2018:barren}
J.~R. McClean, S.~Boixo, V.~N. Smelyanskiy, R.~Babbush, and H.~Neven, ``{Barren
  Plateaus in Quantum Neural Network Training Landscapes},'' \emph{{Nature
  Communications}}, vol.~9, no.~1, p. 4812, 2018,
  \href{https://doi.org/10.1038/s41467-018-07090-4}{doi:10.1038/s41467-018-07090-4}.

\bibitem{holmes:2022:prx}
Z.~Holmes, K.~Sharma, M.~Cerezo, and P.~J. Coles, ``{Connecting Ansatz
  Expressibility to Gradient Magnitudes and Barren Plateaus},'' \emph{PRX
  Quantum}, vol.~3, p. 010313, Jan 2022,
  \href{https://doi.org/10.1103/PRXQuantum.3.010313}{doi:10.1103/PRXQuantum.3.010313}.

\bibitem{cerezo:2021:natcomm}
M.~Cerezo, A.~Sone, T.~Volkoff, L.~Cincio, and P.~J. Coles, ``{Cost Function
  Dependent Barren Plateaus in Shallow Parametrized Quantum Circuits},''
  \emph{Nature Communications}, vol.~12, no.~1, p. 1791, Mar. 2021,
  \href{https://doi.org/10.1038/s41467-021-21728-w}{doi:10.1038/s41467-021-21728-w}.

\bibitem{larocca:2025:natreview}
M.~Larocca, S.~Thanasilp, S.~Wang, K.~Sharma, J.~Biamonte, P.~J. Coles,
  L.~Cincio, J.~R. McClean, Z.~Holmes, and M.~Cerezo, ``{Barren Plateaus in
  Variational Quantum Computing},'' \emph{Nature Reviews Physics}, vol.~7,
  no.~4, pp. 174--189, Apr 2025,
  \href{https://doi.org/10.1038/s42254-025-00813-9}{doi:10.1038/s42254-025-00813-9}.

\bibitem{wang:2021:natcomm}
S.~Wang, E.~Fontana, M.~Cerezo, K.~Sharma, A.~Sone, L.~Cincio, and P.~J. Coles,
  ``{Noise-Induced Barren Plateaus in Variational Quantum Algorithms},''
  \emph{Nature Communications}, vol.~12, no.~1, p. 6961, Nov 2021,
  \href{https://doi.org/10.1038/s41467-021-27045-6}{doi:10.1038/s41467-021-27045-6}.

\bibitem{larocca:2022:quantum}
M.~Larocca, P.~Czarnik, K.~Sharma, G.~Muraleedharan, P.~J. Coles, and
  M.~Cerezo, ``{Diagnosing Barren Plateaus with Tools from Quantum Optimal
  Control},'' \emph{Quantum}, vol.~6, p. 824, Sep. 2022,
  \href{https://doi.org/10.22331/q-2022-09-29-824}{doi:10.22331/q-2022-09-29-824}.

\bibitem{fontana:2024:natcomm}
E.~Fontana, D.~Herman, S.~Chakrabarti, N.~Kumar, R.~Yalovetzky, J.~Heredge,
  S.~H. Sureshbabu, and M.~Pistoia, ``{Characterizing Barren Plateaus in
  Quantum Ans{\"a}tze with the Adjoint Representation},'' \emph{Nature
  Communications}, vol.~15, no.~1, p. 7171, Aug 2024,
  \href{https://doi.org/10.1038/s41467-024-49910-w}{doi:10.1038/s41467-024-49910-w}.

\bibitem{ragone:2024:natcomm}
M.~Ragone, B.~N. Bakalov, F.~Sauvage, A.~F. Kemper, C.~Ortiz~Marrero,
  M.~Larocca, and M.~Cerezo, ``{A Lie Algebraic Theory of Barren Plateaus for
  Deep Parameterized Quantum Circuits},'' \emph{Nature Communications},
  vol.~15, no.~1, p. 7172, Aug 2024,
  \href{https://doi.org/10.1038/s41467-024-49909-3}{doi:10.1038/s41467-024-49909-3}.

\bibitem{zhao:2021:quantum}
C.~Zhao and X.-S. Gao, ``{Analyzing the Barren Plateau Phenomenon in Training
  Quantum Neural Networks with the ZX-Calculus},'' \emph{Quantum}, vol.~5, p.
  466, Jun. 2021,
  \href{https://doi.org/10.22331/q-2021-06-04-466}{doi:10.22331/q-2021-06-04-466}.

\bibitem{li:2026:npj}
K.~Li, Y.~Wang, P.~Gao, and S.~Zheng, ``{Learning Parameterized Quantum
  Circuits with Quantum Gradient},'' \emph{npj Quantum Information}, vol.~12,
  p.~59, Feb 2026,
  \href{https://doi.org/10.1038/s41534-025-01179-7}{doi:10.1038/s41534-025-01179-7}.

\bibitem{puig:2025:prx}
R.~Puig, M.~Drudis, S.~Thanasilp, and Z.~Holmes, ``{Variational Quantum
  Simulation: A Case Study for Understanding Warm Starts},'' \emph{PRX
  Quantum}, vol.~6, p. 010317, Jan 2025,
  \href{https://doi.org/10.1103/PRXQuantum.6.010317}{doi:10.1103/PRXQuantum.6.010317}.

\bibitem{zunkovic:2026:adiabatic}
\BIBentryALTinterwordspacing
B.~Žunkovič, M.~Ballarin, L.~Wright, and M.~Lubasch, ``{Scalable,
  Self-verifying Variational Quantum Eigensolver using Adiabatic Warm
  Starts},'' 2026. [Online]. Available: \url{https://arxiv.org/abs/2602.17612}
\BIBentrySTDinterwordspacing

\bibitem{kandala:2017:hea}
A.~Kandala, A.~Mezzacapo, K.~Temme, M.~Takita, M.~Brink, J.~M. Chow, and J.~M.
  Gambetta, ``{Hardware-Efficient Variational Quantum Eigensolver for Small
  Molecules and Quantum Magnets},'' \emph{Nature}, vol. 549, no. 7671, pp.
  242--246, 2017.

\bibitem{wecker:2015:hva}
D.~Wecker, M.~B. Hastings, and M.~Troyer, ``{Progress towards Practical Quantum
  Variational Algorithms},'' \emph{Phys. Rev. A}, vol.~92, p. 042303, Oct 2015.

\bibitem{park:2024:quantum}
C.-Y. Park and N.~Killoran, ``{Hamiltonian Variational Ansatz without Barren
  Plateaus},'' \emph{Quantum}, vol.~8, p. 1239, Feb. 2024,
  \href{https://doi.org/10.22331/q-2024-02-01-1239}{doi:10.22331/q-2024-02-01-1239}.

\bibitem{schuld:2019:pra}
M.~Schuld, V.~Bergholm, C.~Gogolin, J.~Izaac, and N.~Killoran, ``{Evaluating
  Analytic Gradients on Quantum Hardware},'' \emph{Physical Review A}, vol.~99,
  no.~3, Mar. 2019.

\end{thebibliography}
\end{document}